\documentclass[aps,pre,floats,twocolumn,showpacs,reprint,
  superscriptaddress]{revtex4-1}

\usepackage[latin1]{inputenc}
\usepackage[british]{babel}
\usepackage{graphicx,epsfig}
\usepackage{graphics,dcolumn,bm,fleqn,epic,eepic,float}
\usepackage{amssymb,amsmath,multirow,rotate,color}
\usepackage{subfigure}
\usepackage{xcolor}

\newcommand{\avg}[1]{\langle #1 \rangle}
\newcommand{\ud}{\,\mathrm{d}}

\def\Erdos{Erd\"os}

\begin{document}
\title{Characteristic times of biased random walks on complex networks}

\author{Moreno Bonaventura}
\affiliation{School of Mathematical Sciences, Queen Mary University of
  London, Mile End Road, E1 4NS, London (UK)}
\affiliation{School of Business and Management, Queen Mary University
  of London, Mile End Road, E1 4NS, London (UK)}

\author{Vincenzo Nicosia}
\affiliation{School of Mathematical Sciences, Queen Mary University of
  London, Mile End Road, E1 4NS, London (UK)}

\author{Vito Latora}
\affiliation{School of Mathematical Sciences, Queen Mary University of
  London, Mile End Road, E1 4NS, London (UK)}
\affiliation{Dipartimento di Fisica e Astronomia, Universit\`a di 
Catania and INFN, 95123 Catania, Italy}

\begin{abstract}
We consider degree-biased random walkers whose probability to move
from a node to one of its neighbors of degree $k$ is proportional to
$k^{\alpha}$, where $\alpha$ is a tuning parameter.  We study both
numerically and analytically three types of characteristic times,
namely: {\em i)} the time the walker needs to come back to the
starting node, {\em ii)} the time it takes to visit a given node for
the first time, and {\em iii)} the time it takes to visit all the
nodes of the network. We consider a large data set of real-world
networks and we show that the value of $\alpha$ which minimizes the
three characteristic times is different from the value $\alpha_{\rm
  min}=-1$ analytically found for uncorrelated networks in the
mean-field approximation. In addition to this, we found that
assortative networks have preferentially a value of $\alpha_{\rm min}$
in the range $[-1,-0.5]$, while disassortative networks have
$\alpha_{\rm min}$ in the range $[-0.5, 0]$. We derive an analytical
relation between the degree correlation exponent $\nu$ and the optimal
bias value $\alpha_{\rm min}$, which works well for real-world
assortative networks. When only local information is available,
degree-biased random walks can guarantee smaller characteristic times
than the classical unbiased random walks, by means of an appropriate
tuning of the motion bias.
\end{abstract}


\maketitle


In the last decade or so the quantitative analysis of networks having
different origin and function, including social networks, the human
brain, the Internet, the World Wide Web, has revealed that all these
systems exhibit comparable structural properties at different scales,
and are more similar to each other than
expected~\cite{Newman2003rev,Boccaletti2006}.  It has been found that
the structural complexity of networks from the real world usually has
a significant impact on the dynamical processes occurring over them,
including opinion dynamics~\cite{castellano_09}, 
epidemics~\cite{Vespignani2001} and synchronization~\cite{Arenas2008}.

Random walks are the simplest way to explore a network, and are one of
the most widely studied class of processes on complex
networks~\cite{Noh2004,Yang2005}.  Different kinds of random walks
have been used to implement efficient local search
strategies~\cite{Rosvall2005,ghoshal07}, and also to reveal the
presence of hierarchies and network
communities~\cite{Zhou2003,Rosvall2008}.
Particular attention has been devoted to the study of the
{\em characteristic times} associated to random walks, such as the {\em mean
return times}, or the {\em mean first passage times},
respectively the average time the walker takes to come back to the starting
node or to hit a given node~\cite{Redner2001}.  Such characteristic
times can be determined analytically for random walks on regular
lattices~\cite{Hughes}, but their calculation for graphs with 
heterogeneous structures is still the object of active
research~\cite{zhang2013,yuanlin-zhang2013}. Recent results include the derivation of
analytic expressions for the characteristic times of unbiased random
walks on \Erdos--R\'enyi random graphs~\cite{Sood2005}, on 
fractal networks~\cite{Condamin2007,Agliari2008,Lin2010,Meyer2012} and
on particular classes of scale-free graphs~\cite{Agliari2009}. To
date, only approximate solutions are available for random walks on
real networks~\cite{Tejedor2009,Lau2010,Hwang2012,Baronchelli2006}.

A class of random walks which is particularly interesting to consider
on heterogeneous networks is that of {\em degree-biased random
  walks}. In a degree-biased random walk, the probability to move from
a given node to one of its neighbors, of degree $k$, is proportional
to $k^{\alpha}$, where $\alpha$ is a tuning parameter.  According to
the sign of the bias parameter $\alpha$, the walkers preferentially
move either towards hubs or towards poorly connected
nodes~\cite{Gomez-Gardenes2008}.  Biased random walks have been
recently employed for community detection~\cite{Zlatic2010} and to
define new centrality measures~\cite{Lee2009,Delvenne2011}.
Furthermore, analytical results on the characteristic times of
degree-biased random walks have been obtained for specific classes of
random graphs in the mean-field approximation~\cite{Fronczak2009}.
However, the structure of real networks is far from being random, and
several empirical evidences suggest that the presence of degree-degree
correlations can affect the dynamics of the walk~\cite{Baronchelli2010}.
For instance, the authors of Ref.~\cite{Gomez-Gardenes2008} have shown
that the value of entropy rate of biased random walks on real
correlated networks substantially deviates from the prediction for the
corresponding randomized graphs. Similarly, more recent works show
that degree-biased random walks can approximate maximally entropic
walks, but the quality of such approximation depends again on
degree-degree correlations~\cite{Burda2009,Sinatra2011}.

In this Article we study, both numerically and analytically, three
types of characteristic times for biased random walks, namely mean
return times (MRT), mean first passage times (MFPT), and mean coverage
times (MCT).  We consider different synthetic graphs and a large data
set of social, biological and technological complex networks from the
real world, and we study the effect of the bias parameter $\alpha$ on
the characteristic times of the walk, focusing on the values
$\alpha_{\rm min}$ that guarantee minimal return, first passage and
coverage times.  Our main result is that the characteristic times of
biased random walks on real-world networks sensibly deviate from those
observed in uncorrelated graphs.  In particular, we prove analytically
that the minimum MRT in \Erdos--R\'enyi and scale-free random graphs
is always obtained for $\alpha_{\rm min}=-1$, while we show through
numerical simulations that the minimum MRT in real-world networks is
obtained for values of $\alpha$ that significantly deviate from $-1$.
We find that the value $\alpha_{\rm min}$ depends on the presence and
sign of degree-degree correlations in the network, and in particular
that for assortative networks $-1<\alpha_{\rm min}<-0.5$, while for
disassortative networks $-0.5<\alpha_{\rm min}<0$.  We
  show that in the case of networks in which the expected degree
  $k_{nn}(k)$ of the first neighbors of a node with degree $k$ is a
  power law, i.e. when $k_{nn}(k) \sim k^{\nu}$ as observed in many
  real-world networks, it is possible to derive an approximate
  relation between the optimal bias value $\alpha_{\rm min}$ and the
  exponent $\nu$. This approximation works well for assortative
  networks in which, for any given value of $\nu$, the predicted
  optimal value of $\alpha_{\rm min}$ is close to the real optimum. We
  also analyze the MRT for nodes of a given degree class $k$, and we
  derive a closed form, valid for uncorrelated scale-free graphs, to
  calculate the value of the bias $\alpha_{\rm min}(k)$ which
  minimizes the MRT for nodes of degree $k$.  We also discuss the
  results found for MFPT and MCT, which suggest that the optimal value
  of $\alpha$ for MRT on a given network is a quite accurate
  approximation for the values of $\alpha$ which optimize the MFPT and
  the MCT on the same network.

The paper is organized as follows. In Section~\ref{section:rw} we
introduce degree-biased random walks and we provide the definitions of
return, first passage and coverage times. In Section~\ref{section:mrt}
we study how the MRT depends on the value of the bias parameter
$\alpha$, and we compare the analytical predictions of characteristic
times, which assume the absence of degree correlations, with the
numerical results obtained on a large data set of real-world
networks. We also investigate the dependence of the MRT on the degree
of the starting node. In Section~\ref{section:mfpt} and
Section~\ref{section:mct} we study, respectively, the behavior of
MFPT and MCT on real-world networks, and we show that the relation
between the sign of degree-degree correlations in a graph and the
dynamics of the walkers on the graph are indeed similar to those found
for MRT. In Section ~\ref{section:discussion} we
  provide a more detailed discussion of the results presented in the
  paper, we derive an analytical relation between $\nu$ and
  $\alpha_{\rm min}$, and we indicate possible applications to several
  problems connected with characteristic times of biased random walks.
  Finally, in Section~\ref{section:conclusions} we draw some
  conclusions and we suggest possible future directions of research in
  this field.

\section{Degree-Biased Random Walks}
\label{section:rw}
Let us consider an undirected and unweighted graph $G=(V,E)$ with
$N=|V|$ nodes and $K=|E|$ edges. Denote as $A$ the adjacency matrix of
graph $G$, i.e. the symmetric $N \times N$ matrix whose entry $a_{ij}$
is equal to $1$ if an edge exists between node $i$ and $j$, and is $0$
otherwise.  We consider the following dynamical process occurring on
the graph: a walker that at each time step moves from a node to one of
its neighbors with a probability proportional to the $\alpha$-power
of the degree of the target node.
The process corresponds to a discrete-time 
Markov chain \cite{norris1998} on the state space $V$ defined by the transition 
matrix $\Pi$, whose each entry
$\pi_{ji}$ is equal to the probability for a walker on node $i$ to
jump to one of its neighbors $j$, and reads:
\begin{equation}  
    \pi_{ji}  =  \frac{  a_{ij}  k_j^{\alpha}} {\sum_l a_{il} k_l^{\alpha}}
  \label{eq:biased_trans_matrix}
\end{equation}
The exponent $\alpha$ is the control parameter that allows to tune the
dependence of the process on the node degree.  When $\alpha >0$ the
random motion is biased towards \mbox{high-degree} nodes (hubs), while
when $\alpha <0$ the walkers move with higher probability to neighbors
with low degree. When $\alpha=0$ the common (unbiased) random walk is
recovered.  The fundamental quantity to describe a random walk is the
occupation probability distribution $p_i(t)$.  Being $p_i(t)$ the
probability that a walker is at node $i$ at time $t$, then the
probability $p_j(t+1)$ of being at node $j$ at time $t+1$ is given by:
\begin{equation}
p_j (t+1) = \sum_i \pi_{ji} p_i(t)
\label{eq:pievolution}
\end{equation}
or in vector notation: $\bm{p}(t+1)=\Pi \bm{p}(t)$.  A fixed point
solution $\bm{p^{*}}$ of the latter equation, such that
$\bm{p^{*}}=\Pi \bm{p^{*}}$, is called {\em stationary distribution}.
If the transition matrix $\Pi$ is primitive, i.e. if the graph is
connected and contains at least one odd cycle, the Perron-Frobenius
theorem guarantees that $\bm{p^*}$ always exists, is unique, and
\begin{equation*}
\lim_{t \to \infty} \Pi^{t} \, \bm{p}(0) = \bm{p^*}
\,\,\,\,\,\,\, \forall \,\, \bm{p}(0)
\end{equation*}
i.e. all initial occupation probability distributions $\bm{p}(0)$ converge 
to the stationary distribution $\bm{p^*}$ \cite{Cover1991}. 
In particular, the stationary distribution associated to the
transition matrix (\ref{eq:biased_trans_matrix}) of a degree-biased
random walk is~\cite{Gomez-Gardenes2008}:
\begin{equation}
p^{*}_{i} = \frac{ c_i k_i^{\alpha} } {\sum_{\ell} c_{\ell}
  k^{\alpha}_{\ell} }, {\rm ~~~~~~} c_i = \sum_j a_{ij} k_j^{\alpha}
\label{jesus_prob}
\end{equation} 
When $\alpha=0$, Eq.~(\ref{jesus_prob}) reduces to:
\begin{equation}
p_i^* = \frac{k_i}{2K} 
\label{eq:stationary_unbiased}
\end{equation}
which states that for unbiased random walks the number of walkers at a
node $i$ is proportional to the degree $k_i$, so that the dynamic
process is completely characterized by the degree sequence of the
graph. Conversely, when $\alpha \neq 0$, the stationary distribution
$p_i^*$ does not depend only on the degree $k_i$ but also on the
degrees of the first neighbors of node $i$, through the coefficient
$c_i$.  The stationary probability distribution $\bm{p}^*$ is
therefore sensitive to the degree sequence and also to the presence of
degree-degree correlations in the network. It is interesting to notice
that the majority of real-world networks exhibit degree-degree
correlations, meaning that their nodes are found to be preferentially
connected with other nodes having either similar or dissimilar
degree~\cite{Pastor-Satorras2001,Newman2002,Newman2003}. Consequently,
in these networks the stationary probability distribution $\bm{p^*}$
can sensibly deviate from that observed on a random graph having the
same degree distribution and no degree-degree correlations.
Degree-degree correlations are fully described by the joint
probability $P(k,k')$, that represents the likelihood that nodes with
degree $k$ and $k'$ are connected through an edge, or equivalently by
the conditional probability distribution $P(k'|k)$, which represents
the probability that a node of a given degree $k$ is connected to a
node of degree $k'$. The type of correlations is usually characterized
by the average degree $k_{nn} (k)$ of the nearest neighbors of nodes
with degree $k$. This can be written in terms of the conditional
probability distribution $P(k'|k)$ as~\cite{Pastor-Satorras2001}:
\begin{equation*}
k_{nn} (k) = \sum_{k'} k' P(k'|k)
\end{equation*}
Networks are called assortative when $k_{nn}$ is an increasing
function of $k$ and disassortative when $k_{nn}$ is a decreasing
function of $k$~\cite{Pastor-Satorras2001}.
In many real-world networks the nearest neighbors average degree is
found to be a power-law function of $k$, i.e. $k_{nn}(k) \sim
k^{\nu}$, so that the exponent $\nu$ ---often called degree
correlation exponent--- can be used to quantitatively characterize
degree correlations. A positive exponent $\nu>0$ indicates assortative
correlation while a negative value $\nu<0$ indicates disassortative
ones.

In this paper we are interested in the typical times of degree-biased
random walks.  In particular, assuming that a walker is at node $i$ at
time $t=0$ and moves according to Eq.~(\ref{eq:biased_trans_matrix}),
we consider the expected time that the random walker needs to:

\begin{itemize}
\item come back to node $i$ for the first time, referred to as Mean
  Return Time (MRT) and denoted as $r_i$,

\item reach a node $j$ ($j \ne i$) for the first time, referred to as
  Mean First Passage Time (MFPT) and denoted as $t_{ij}$,

\item visit all nodes in the network at least once, referred to as
  Mean Coverage Time (MCT) and denoted as $c_i$.
\end{itemize}

In the following sections we explore how the three characteristic
times defined above are affected by the bias in the random walk. In
particular we will focus on the value of the bias parameter $\alpha$
which respectively minimizes MRT, MFPT and MCT. We use a data set
consisting of many assortative and disassortative medium-to-large
sized real-world networks, and we will show how degree biased random
walks can highlight assortativity or disassortativity from a 
dynamical point of view.

\section{Mean Return Time}
\label{section:mrt}
It is possible to prove that the mean return time $r_i$ of a random
walk on a graph is equivalent to the inverse of the stationary
distribution of the walk~\cite{grinstead}:
\begin{equation}
 r_i = 1 / p^*_i
\label{eq:returntime}
\end{equation}
In order to summarize in a single value the typical return time for
the entire network, we define the {\em graph mean return time} $R$ as
the average of $r_i$ over all nodes:
\begin{equation}
R = \langle r_i \rangle = \frac{1}{N} \sum_{i=1}^N r_i
\label{eq:net_meanRT}
\end{equation}

\begin{figure}[!t]
  \centering
  \includegraphics[width=3.4in]{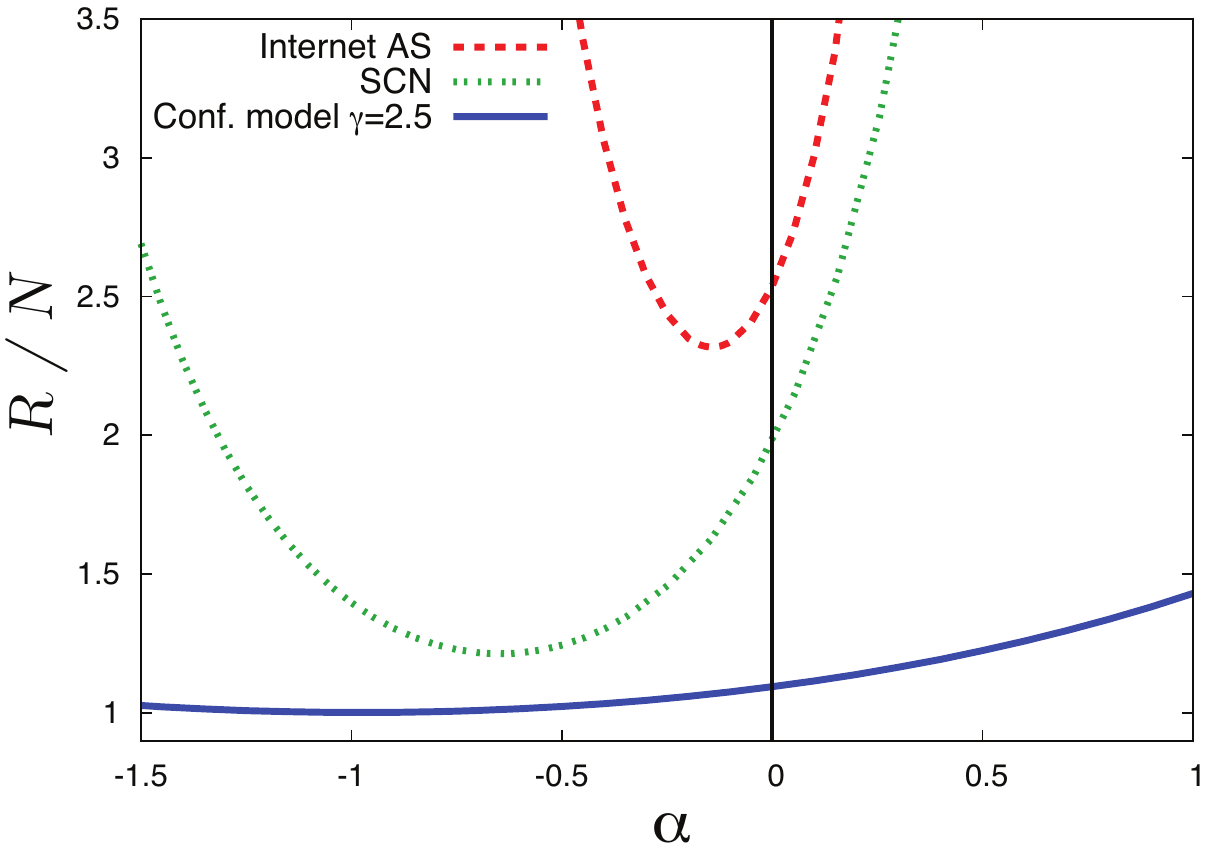}
  \caption{(color online) The graph mean return time $R$
    rescaled by the number of nodes $N$ as a function of
    $\alpha$ for InternetAS (dashed red line), SCN (dotted green line)
    and an uncorrelated scale-free network (solid blue line). Due to
    the presence of correlations, $R/N$ is a much narrower function of
    $\alpha$ in real-world networks than in synthetic networks,
    suggesting that mean-field approximations can adequately describe
    the dynamics of biased random walks only for uncorrelated graphs.}
\label{fig:mrt_of_assort_disassor}
\end{figure}

\textit{Empirical evidences. ---} In the case of a degree-biased random
walk, $R$ depends on $\alpha$ because the stationary distribution
depends on $\alpha$ as in Eq.~(\ref{jesus_prob}).
In Fig.~\ref{fig:mrt_of_assort_disassor} we show the graph mean return
time $R$ as a function of $\alpha$ for three networks, namely a
scale-free network with $N=10^4$ nodes, $P(k)\sim k^{-\gamma}$ with
$\gamma=2.5$ and average degree $\avg{k}=46$, constructed by the
configuration model~\cite{bender}, the scientific collaboration
network of scientists in condensed matter (SCN)~\cite{Newman2001},
having $N= 12,722$ nodes and $K=39,967$ edges, and a sample of the
Internet at the Autonomous System level
(InternetAS)~\cite{Pastor-Satorras2001}, having $N=11,174$ and
$K=23,409$ edges. The values of $R$ are rescaled by the network size
$N$.
\begin{figure}[!t]
    \includegraphics[width=3.4in]{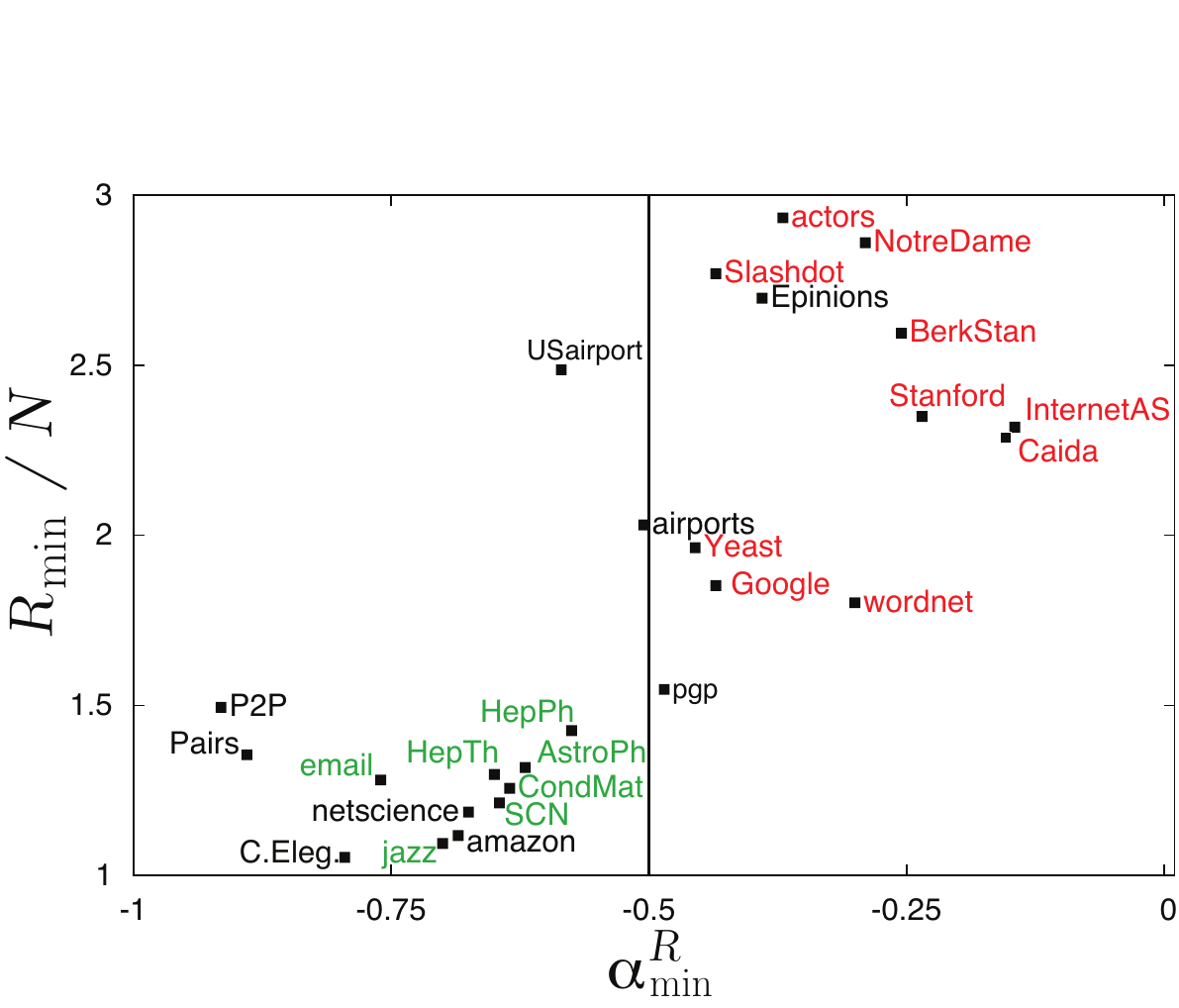}
\caption{(color online) The minimum value of the normalized graph mean
  return time $R_{\rm min}/N$ and the corresponding $\alpha^R_{\rm
    min}$ for all the networks in the considered data set. Assortative
  networks (green labels) have $-1<\alpha^R_{\rm min }<-0.5$ and
  $R_{\rm min}/N \in [1.0, 1.5]$, while for disassortative networks
  (red labels) $-0.5<\alpha^R_{\rm min }<0$ and $R_{\rm min}/N >
  1.5$. Black labels indicate those networks for which the function
  $k_{nn}(k)$ is not a power-law of $k$.}
\label{fig:mrt_rescaled}
\end{figure}

The networks reported in Fig.~\ref{fig:mrt_of_assort_disassor} are
representative of the general behavior observed in the entire data
set. In fact, for all the considered networks $R$ is always a convex
function of $\alpha$, with a single minimum, denoted by $R_{\rm min}$,
observed at a value of $\alpha$ denoted as $\alpha^R_{\rm min}$.  For
the uncorrelated scale-free network we find $\alpha^R_{\rm min} = -1$
and $R_{\rm min} \sim N$. The same result has been found for
\Erdos--R\'enyi random graphs and for other uncorrelated scale-free
networks constructed through linear preferential
attachment~\cite{Albert1999}. As shown in Table~\ref{tab:list_net}, as
the average degree $\langle k \rangle$ of a synthetic network
increases, the corresponding value of $\alpha^R_{\rm min}$ approaches
$-1$. Also the minimum return time becomes progressively more similar
to the size of the network: $R_{\rm min} \sim N$.
These results are in agreement with what has been found in
Ref.~\cite{Fronczak2009}.  We will give an analytical explanation of
the fact that $\alpha^R_{\rm min}=-1$ for uncorrelated networks at 
the end of this Section.

From Fig.~\ref{fig:mrt_of_assort_disassor} it is clear that the
dynamical behavior of biased random walks on real-world networks
considerably deviates from that observed in uncorrelated synthetic
networks. In fact, if a network has degree-degree correlations then
the minimum of $R$ always occurs for values of $\alpha$ larger than
$-1$. In particular, for SCN we have $\alpha_{\rm min}^R\simeq -0.65$
while for InternetAS we have $\alpha_{\rm min}^R\simeq -0.15$ (refer
to Table~\ref{tab:list_net} for the values of $\alpha_{\rm min}^R$ in
each of the real-world networks considered in this study).
As we see in the Figure, the value of $R$ in real-world networks is
also highly sensitive to the value of $\alpha$, and $R_{\rm min}$ can
be considerably lower than the value of MRT corresponding to an
unbiased random walk ($\alpha=0$) on the same network.  For instance,
in SCN the value of $R_{\rm min}$ is about half the value of $R$
corresponding to an unbiased random walk.  This result indicates that
a carefully chosen value of the bias parameter can significantly
reduce the characteristic times of degree-biased random walks.

In Fig.~\ref{fig:mrt_rescaled} we report the values of $R_{\rm min}$
and $\alpha^R_{\rm min}$ for each network in the data set. For those
networks with $\alpha^R_{\rm min} < -0.5$ the minimum value $R_{\rm
  min}$ is only slightly greater than the size of the network $N$,
while the differences are more pronounced in the region $\alpha>
-0.5$.  Notice that all the networks with clear assortative
degree-degree correlations (reported in green) have a value of
$\alpha^R_{\rm min} < -0.5$, while disassortative networks (reported
in red) have $\alpha^R_{\rm min} > -0.5$.
This result indicates that the presence of degree-degree correlations
has a significant impact in the values of $\alpha^R_{\rm min}$, and
consequently on the performance of a biased random walk on a graph in
terms of exploration speed.  

\begin{figure}[!b]
\centering
\includegraphics[width=3.4in]{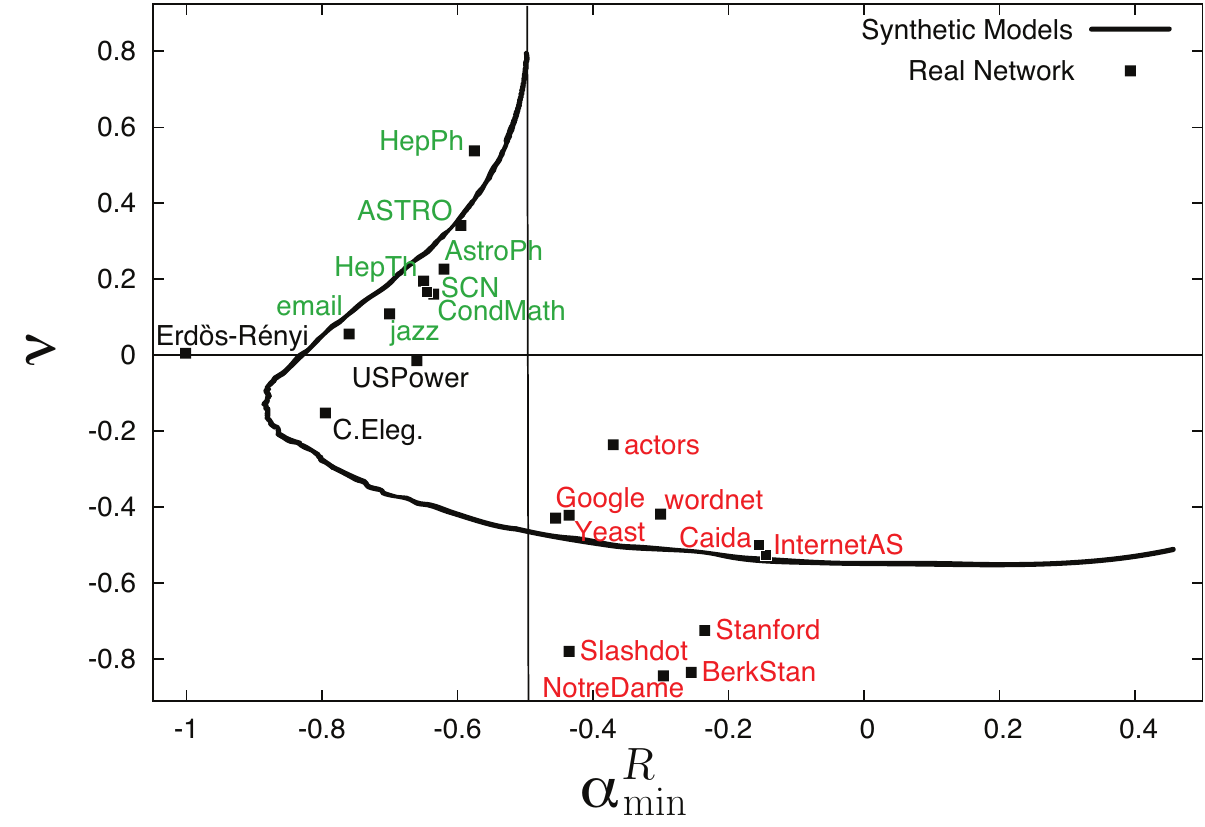}
\caption{(color online) The degree correlation exponent and the value
  $\alpha^{R}_{\rm min}$ that minimizes the graph mean return time in
  real-world networks (black squares) and in synthetic networks with a
  tunable value of $\nu$ (solid black line). Notice that assortative
  networks are confined in $-1<\alpha^{R}_{\rm min}<-0.5$ while almost
  all the disassortative networks lie in the region
  $-0.5<\alpha^{R}_{\rm min}<0$.  For comparison we also report the
  value of $\alpha^{R}_{\rm min}$ for an \Erdos --R\'enyi random graph
  with $N=10^4$ and $\langle k \rangle=40$, which is equal to
  $\alpha=-1$ as predicted by the theory for uncorrelated networks.}
\label{fig:isteresi}
\end{figure}
%


The relation between the degree-correlation exponent $\nu$ and the
value of $\alpha^R_{\rm min}$ is shown in Fig.~\ref{fig:isteresi}.
The values corresponding to real-world networks lie almost exclusively
in the top-left and in the bottom-right quadrants, respectively
corresponding to $(\alpha^R_{\rm min}<-0.5 , \nu > 0 )$ and
$(\alpha^R_{\rm min} > -0.5,\nu < 0)$.  Fig.~\ref{fig:isteresi} shows
very clearly that the value of $\alpha_{\rm min}$ is always in the
interval $[-1.0,-0.5]$ for assortative networks and larger than $-0.5$
for disassortative ones.

To further investigate the special role played by the bias parameter
$\alpha=-0.5$ we have considered a large set of synthetic networks, in
which we tuned the level and sign of degree-degree correlations
through the edge-swapping procedure described in
Ref.~\cite{BrunetSokolov2005}. This procedure, discussed in details in
Appendix, starts from an uncorrelated network and artificially
introduces a prescribed amount of either assortative or disassortative
degree-degree correlations by rewiring the edges of the graph without
modifying the degree sequence. As a result, this algorithm allows to
investigate the relation between the value of $\nu$ and $\alpha^R_{\rm
  min}$ of a network by varying continuously the correlation exponent
$\nu$ while preserving the degree sequence.

The black curve in Fig.~\ref{fig:isteresi} has been obtained by
starting with a configuration model scale-free network with $N=10^4$
nodes, $k_{max}=300$ and $\gamma=3$~\cite{note_cutoff}, and by running
the swapping procedure to introduce assortative or disassortative
correlations. We notice that by performing assortative swaps the value
of $\nu$ increases considerably, while $\alpha^R_{\rm min}$ remains
asymptotically confined below $-0.5$. Conversely, few disassortative
swaps are enough to determine a fast change on $\alpha^R_{\rm min}$, which
enters the region $\alpha>-0.5$ where the majority of real-world
disassortative networks lie.

\begin{figure}[!t]
\centering
\includegraphics[width=3.4in]{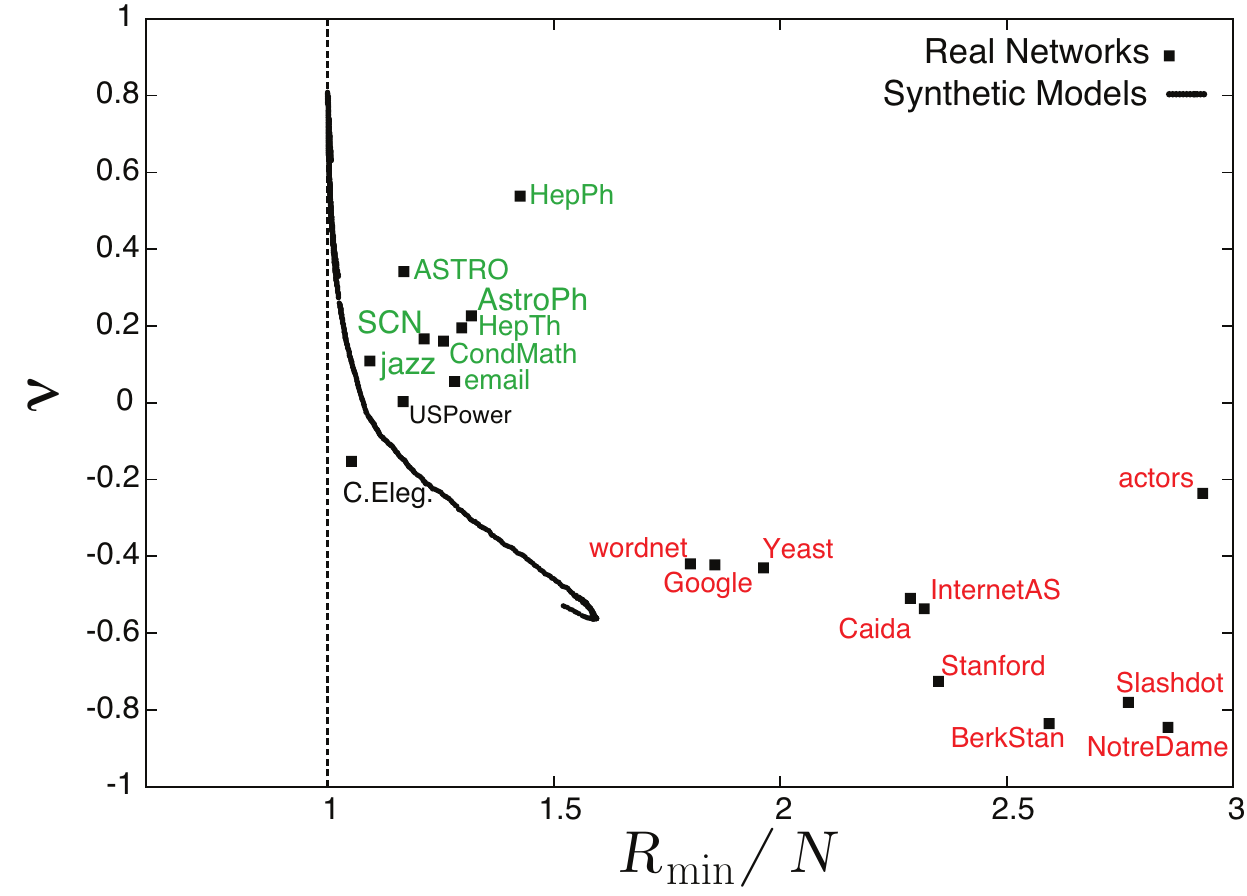}
\caption{(color online) The degree-correlation exponent $\nu$ and the
  normalized graph mean return time $R_{\rm min}/N$ for the same set
  of synthetic networks in Fig.\ref{fig:isteresi} (solid black line)
  and for real-world networks (black squares). Notice that in the
  $\nu-R_{\rm min}/N$ plane all real-world networks lie on the right
  of the curve corresponding to synthetic correlated networks.}
\label{fig:isteresiMRT}
\end{figure}

\mbox{In Fig.~\ref{fig:isteresiMRT}} we report as a solid line the
values of $R_{\rm min}$ as a function of the degree-correlation
exponent $\nu$ for the same set of synthetic networks considered in
Fig.~\ref{fig:isteresi}. Filled squares represent the values obtained
on real-world networks. We observe that $R_{\rm min}/N$ is
considerably larger than $1$ for disassortative networks, while it is
closer to $1$ for assortative networks. Notice that the MRT of the
synthetic network with tunable degree correlations (solid black line)
is consistently smaller than that of any real-world network, with the
only exception of the C.Elegans neural network.

%
%
\textit{Analytical arguments. ---} The numerical analysis of MRT
suggests that, for uncorrelated networks, $\alpha_{\rm min}=-1$, so
that the deviations from this value observed in real-world networks
should be due to the presence of degree-degree correlations. Here, we
provide an analytical proof of the fact that $\alpha_{\rm min}=-1$ for
uncorrelated graphs in the mean-field approximation, and we compare
this analytical prediction with numerical results on real-world
networks.  In the mean-field approximation a graph is described by the
annealed adjacency matrix:
\begin{equation}
{\langle a \rangle }_{ij} = \frac{k_i k_j}{2K}
\label{eq:annealed_adj}
\end{equation}
where the value $\avg{a}_{ij}$ represents the probability to find an
edge connecting node $i$ and node $j$, having degrees $k_i$ and $k_j$,
if the nodes are connected uniformly at random.
In fact, let us imagine a network where each node $i$ has $k_i$ stubs
to be paired with some of the stubs of other nodes. If $K$ is the
total number of links there are $2K$ of such free stubs. Among these
$2K$ stubs, only $k_j$ are incident on node $j$. Therefore, there are
$k_j$ ways a stub of node $i$ can be connected with node $j$ over a
total of $2K$ possible pairings with other nodes.  One obtains the
expression for $\avg{a}_{ij}$ in Eq.~(\ref{eq:annealed_adj}) by
observing that node $i$ has $k_i$ different stubs to connect with one
of the stubs of $j$. If we plug Eq.~(\ref{eq:annealed_adj}) into
Eq.~(\ref{jesus_prob}) we obtain:
\begin{equation}
 p_i^* = \frac { k_i^{\alpha+1} } { N \langle k^{\alpha+1} \rangle}
 \label{eq:pstar_mean_field}
\end{equation}
which gives
\begin{equation}
r_i (k_i) = N \langle k^{\alpha +1} \rangle k_i^{-\alpha-1}
\label{eq:mrt_k_meanfield}
\end{equation}  
and
\begin{equation}
R = N \langle k^{\alpha+1} \rangle \langle k^{-\alpha-1} \rangle
\label{eq:mrt_meanfield}
\end{equation}
in agreement with the result found in Ref.~\cite{Fronczak2009}.

\begin{figure*}[!ht]
  \centering
  \subfigure[]{
    \includegraphics[width=2in]{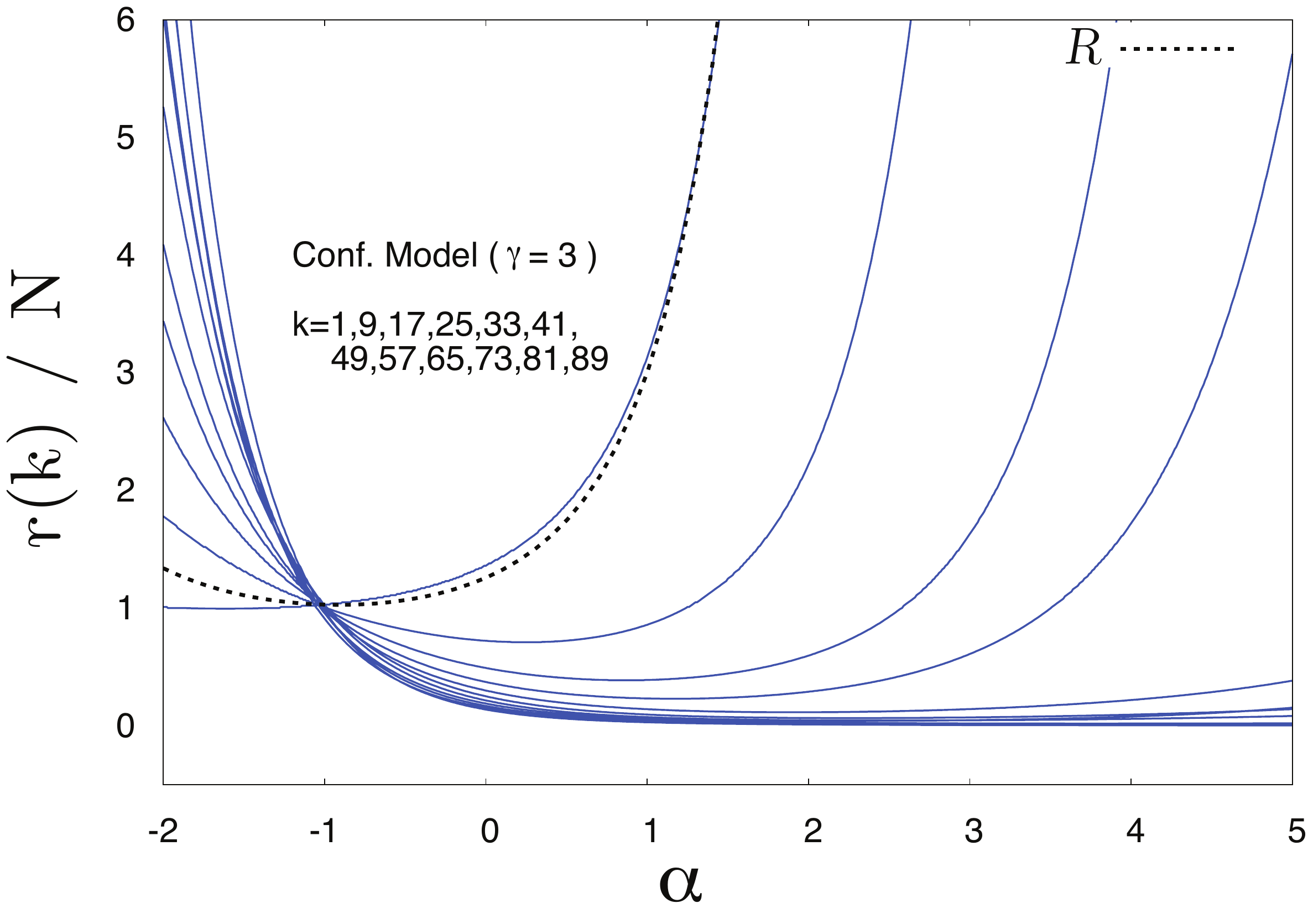}
    \label{fig:overlap_cm}
    }
  \subfigure[]{
       \includegraphics[width=2in]{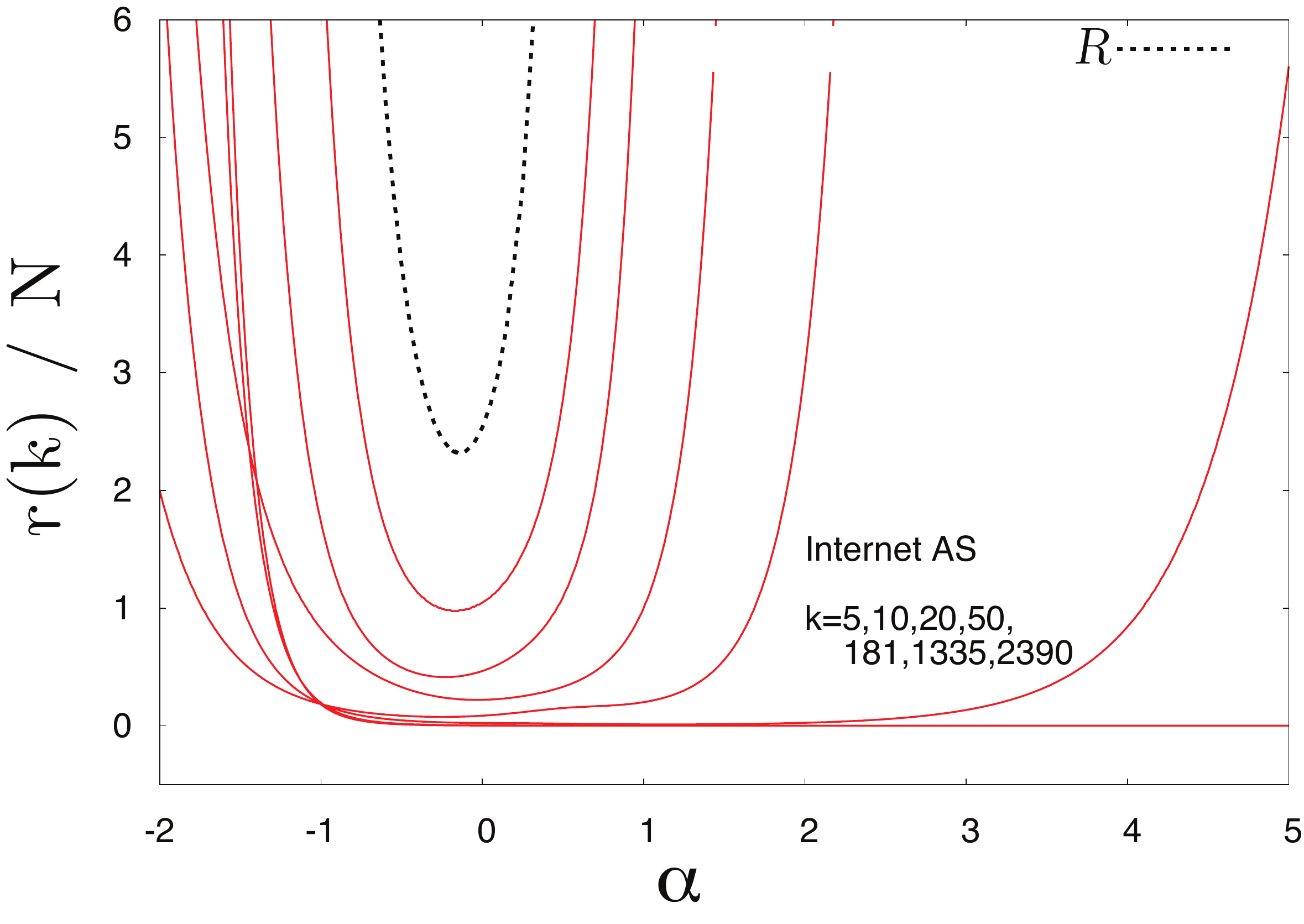}
   \label{fig:overlap_uspower}
   }
  \subfigure[]{
    \includegraphics[width=2in]{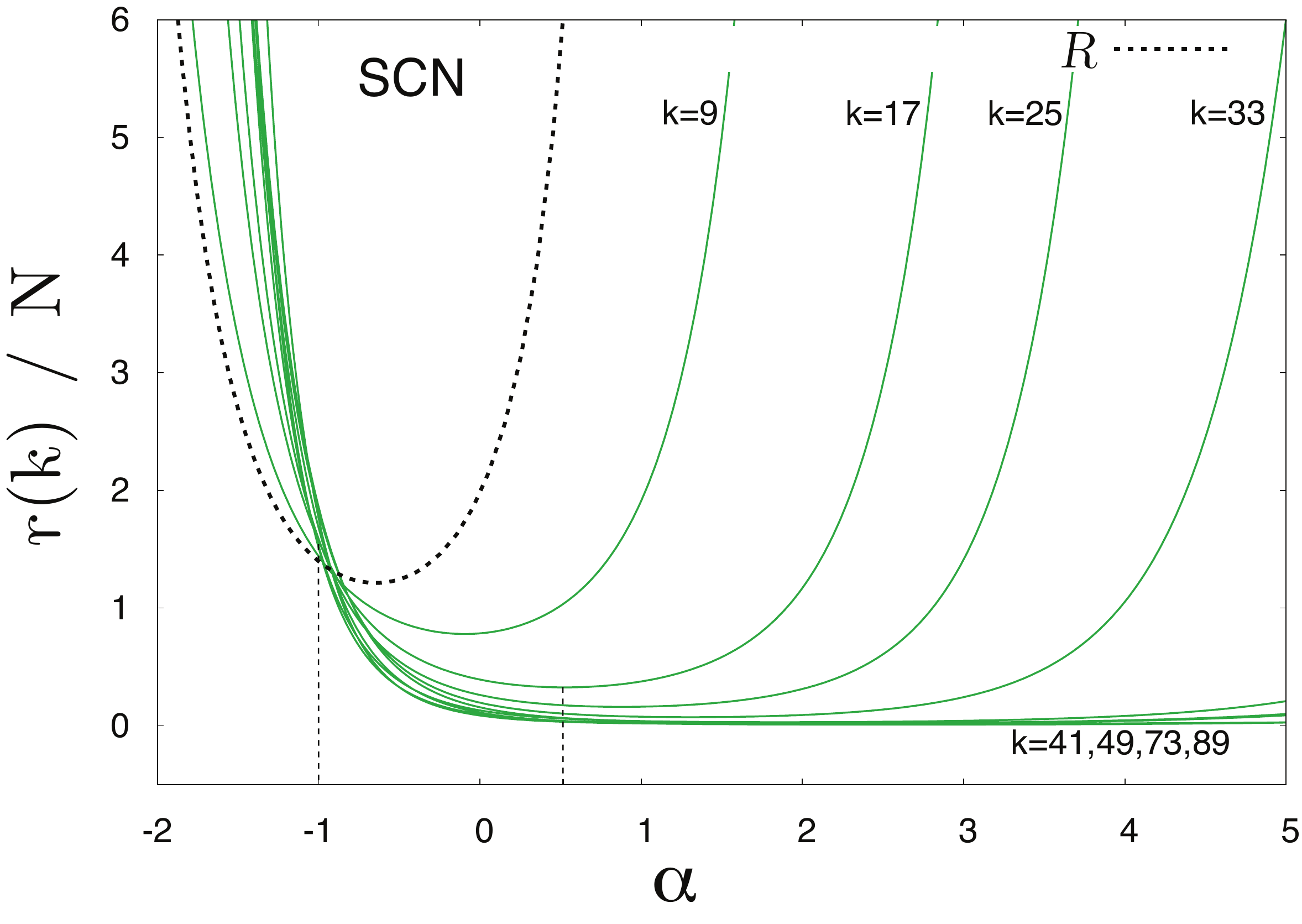}
    \label{fig:overlap_ias}
    }
\\
  \subfigure[]{
    \includegraphics[width=2in]{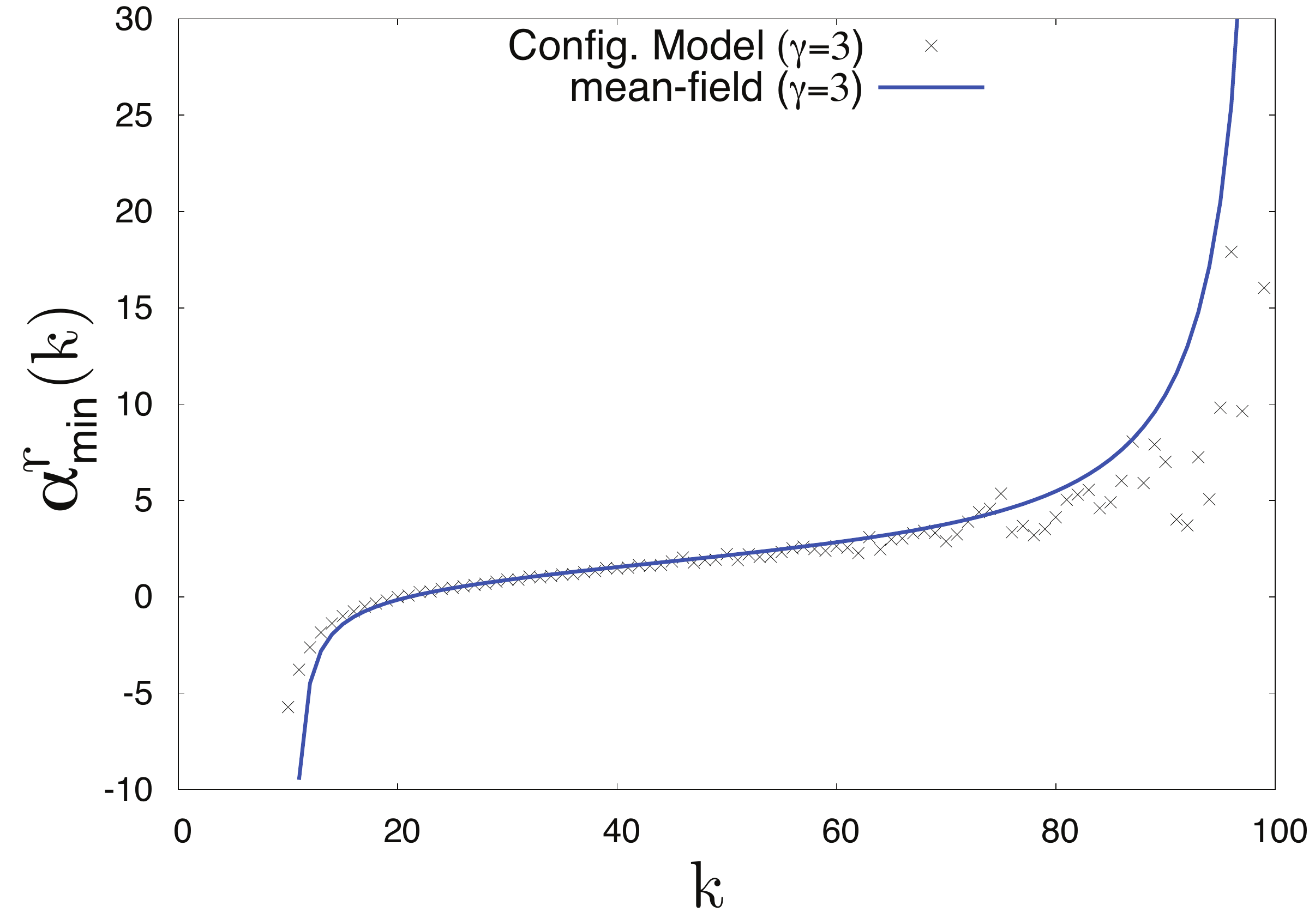}
      \label{fig:cm_amin_of_k}
      }
   \subfigure[]{
      \includegraphics[width=2in]{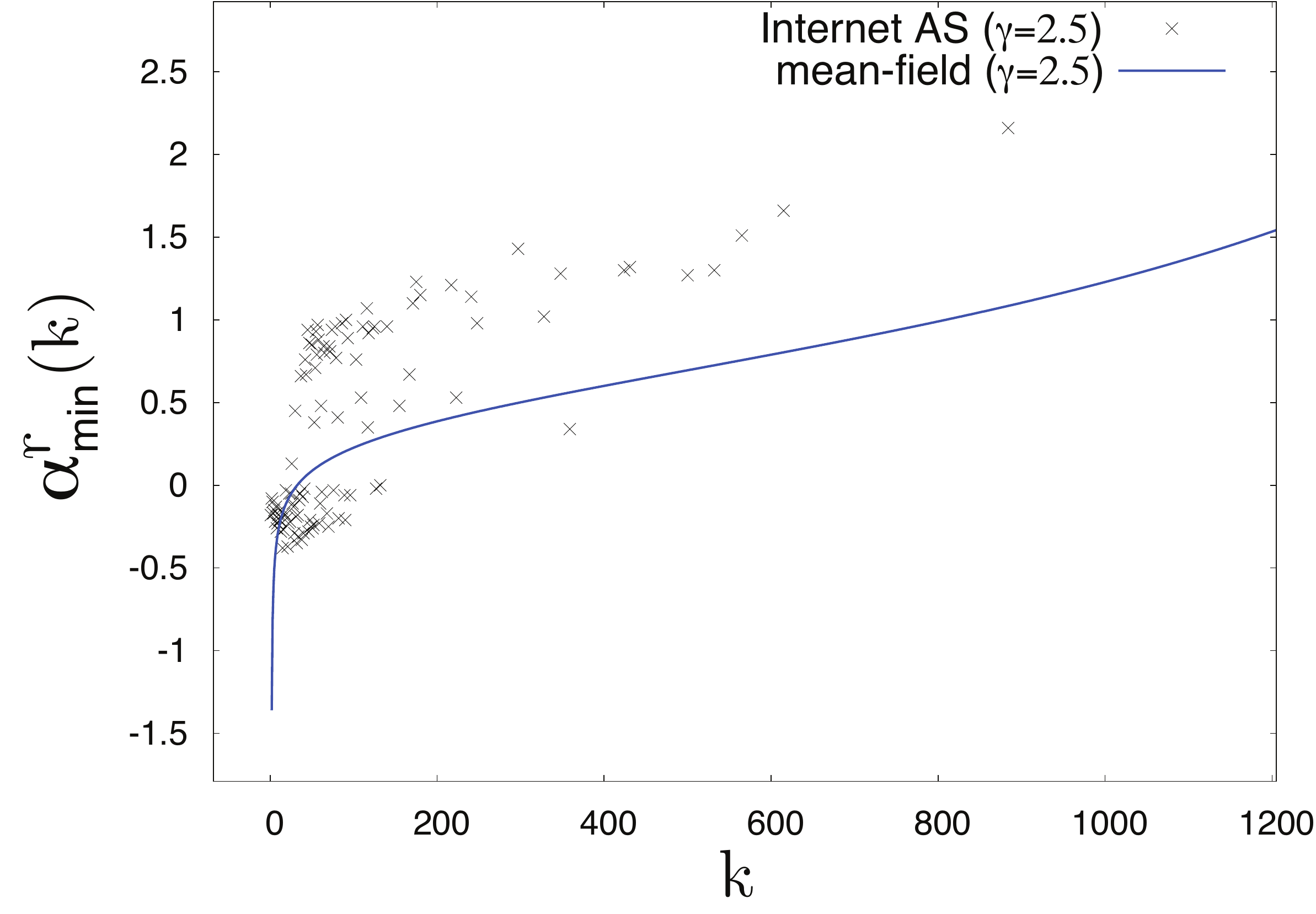}
       \label{fig:amin_di_k_scn}
       }
  \subfigure[]{
    \includegraphics[width=2in]{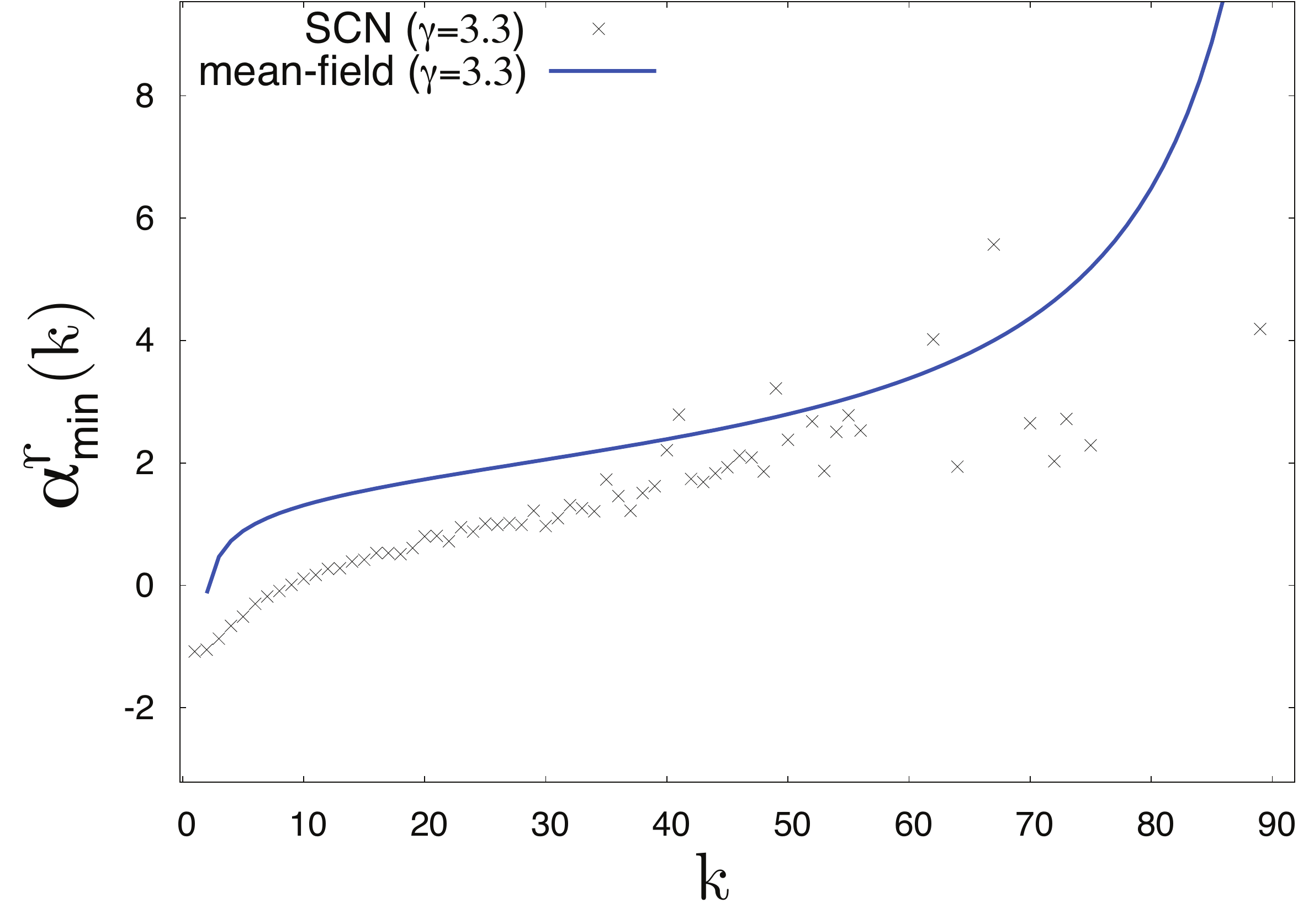}
     \label{fig:amin_di_k_internet}
     }
  \caption{(color online) Top panels: return time $r(k)$ for nodes of
    degree class $k$ as a function of $\alpha$ (solid lines, each
    curve correspond to a value of $k$) respectively for (a) a
    configuration model scale-free graph with $\gamma = 3$ (blue), (b)
    InternetAS (red) and (c) SCN (green). The dotted black line in
    each panel corresponds to the average of $r(k)$ over all degree
    classes. Bottom panels: the value $\alpha^{r}_{\rm min}(k)$ which
    minimizes $r(k)$ as a function of $k$ (dots) for the three
    networks considered in the top panels. The solid blue line is the
    mean-field prediction of Eq.~(\ref{eq:a_min_of_k_cutoff}) where
    $\gamma$ is chosen equal to the exponent of the degree
    distribution of the corresponding network.}
  \label{fig:figure2}
\end{figure*}
It is straightforward to verify that $R = N$ when $\alpha=-1$.
Moreover, one can easily verify that for \Erdos--R\'enyi graphs the
minimum value of $R$ is obtained for $\alpha = \alpha_{\rm
  min}=-1$. In order to see this fact, we replace the average over
nodes $\langle\ldots \rangle$ in Eq. (\ref{eq:mrt_meanfield}) with an
integral over degree classes $\int_1^{\infty} \ldots ~P(k)\, \ud k $.
We denote with $P_{ER}(k)$ the degree distribution of \Erdos--R\'enyi
graphs (this distribution is binomial, and can be approximated by a
Poisson distribution for large $N$).  Differentiating with respect
to $\alpha$ to find the minimum value of $R$ we have:
\begin{equation}
\begin{split}
& 0= \frac{dR}{d\alpha} = \\ & = N \frac{d}{d \alpha} \left[
    \int_1^{\infty} P_{ER}(k) k^{\alpha+1} \, \ud k \int_1^{\infty}
    P_{ER}(z) z^{-\alpha-1} \, \ud z \right] = \\ & = N\int_1^{\infty}
  P_{ER}(k) log(k) k^{\alpha+1} \, \ud k \int_1^{\infty} P_{ER}(z)
  z^{-\alpha-1} \, \ud z \,\, + \\ & - N\int_1^{\infty} P_{ER}(k)
  k^{\alpha+1} \, \ud k \int_1^{\infty} P_{ER}(z) log(z) z^{-\alpha-1} \,
  \ud z.
\end{split} \nonumber
\end{equation}
The latter expression is equal to 0 when $\alpha=-1$, since we have $
k^{\alpha+1} = 1 = z^{-\alpha-1} $ and the last two terms are equal
and opposite in sign.
Analogously we can derive the minimum value of $R$ also for
uncorrelated networks with power-law degree distribution $P(k) \sim
k^{-\gamma}$:
\begin{equation}
\centering
\begin{split}
& R \sim N \int_{1}^{\infty} k^{-\gamma} k^{\alpha +1} dk
  \int_{1}^{\infty} k^{-\gamma} k^{ - \alpha -1} dk = \\ & = N \left[
    \frac{1}{\gamma - \alpha - 2} \right] \left[ \frac{1}{\gamma +
      \alpha} \right]
\end{split}
\label{eq:mrt_meanfield_integral}
\end{equation}

where the integrability conditions are satisfied if $ \alpha $ is in
the range $[-2,0]$, and $ 2 < \gamma < 4 $ which is compatible with
the values of $\gamma$ observed in real-world networks.


\begin{figure*}[!t]
\centering
  \subfigure[]{
  \includegraphics[width=2.24in]{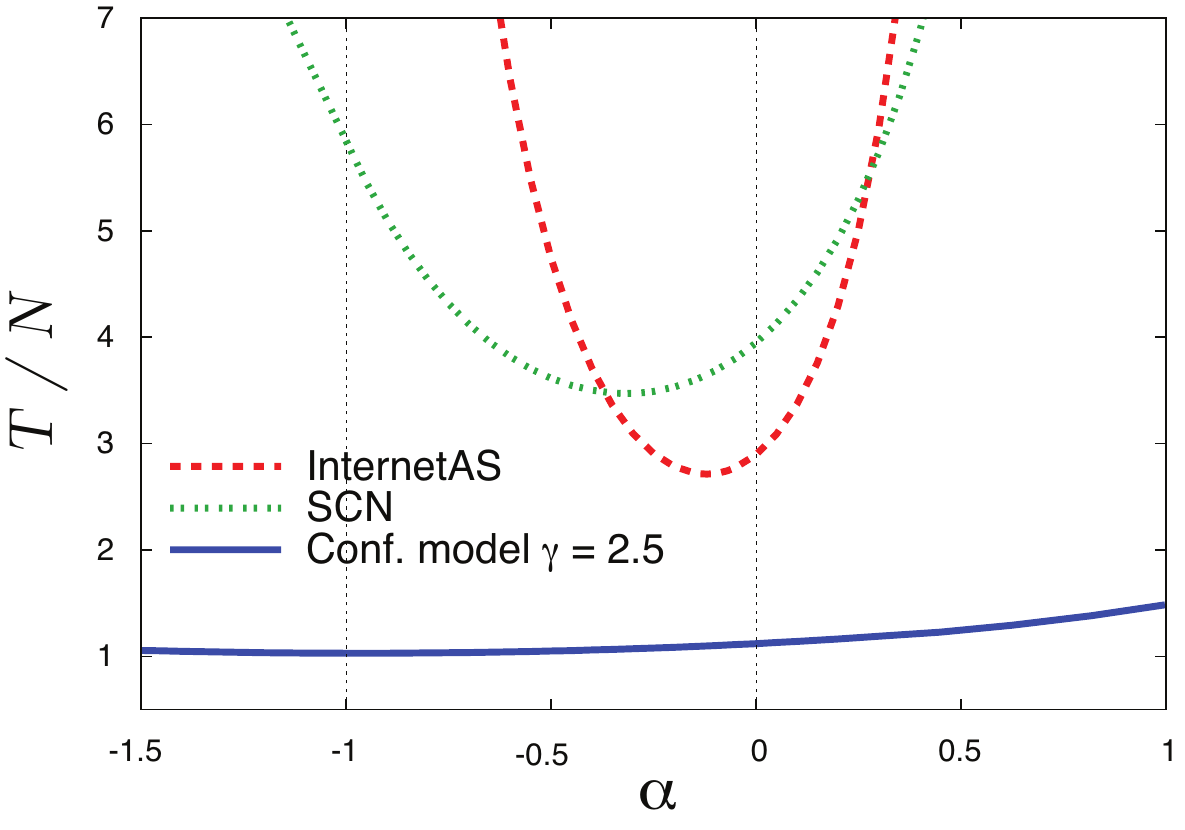}
  \label{fig:mfpt_of_assort_disassor}
   }
  \subfigure[]{
  \includegraphics[width=2.24in]{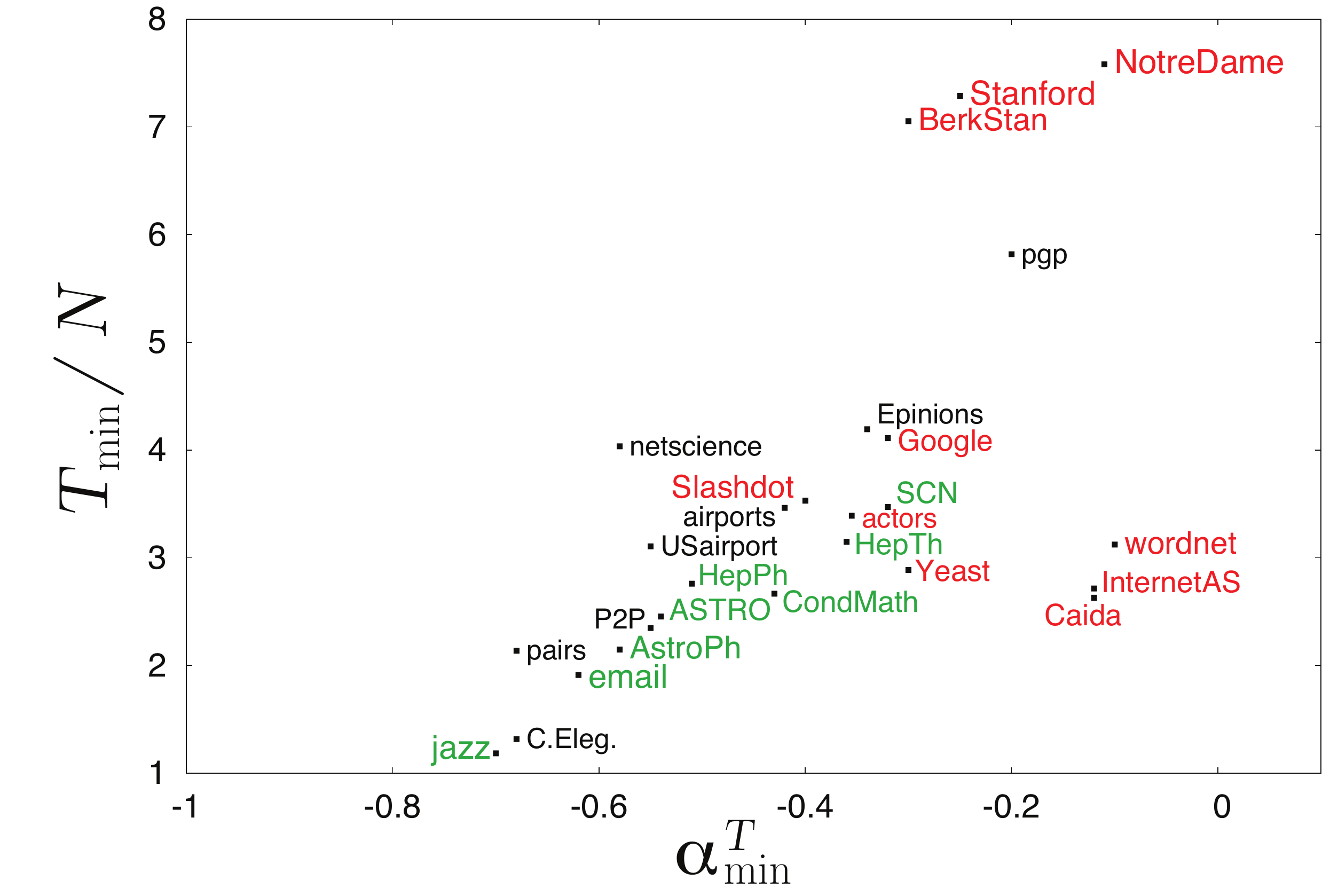}
  \label{fig:mfpt_rescaled}
  }
  \subfigure[]{
  \includegraphics[width=2.24in]{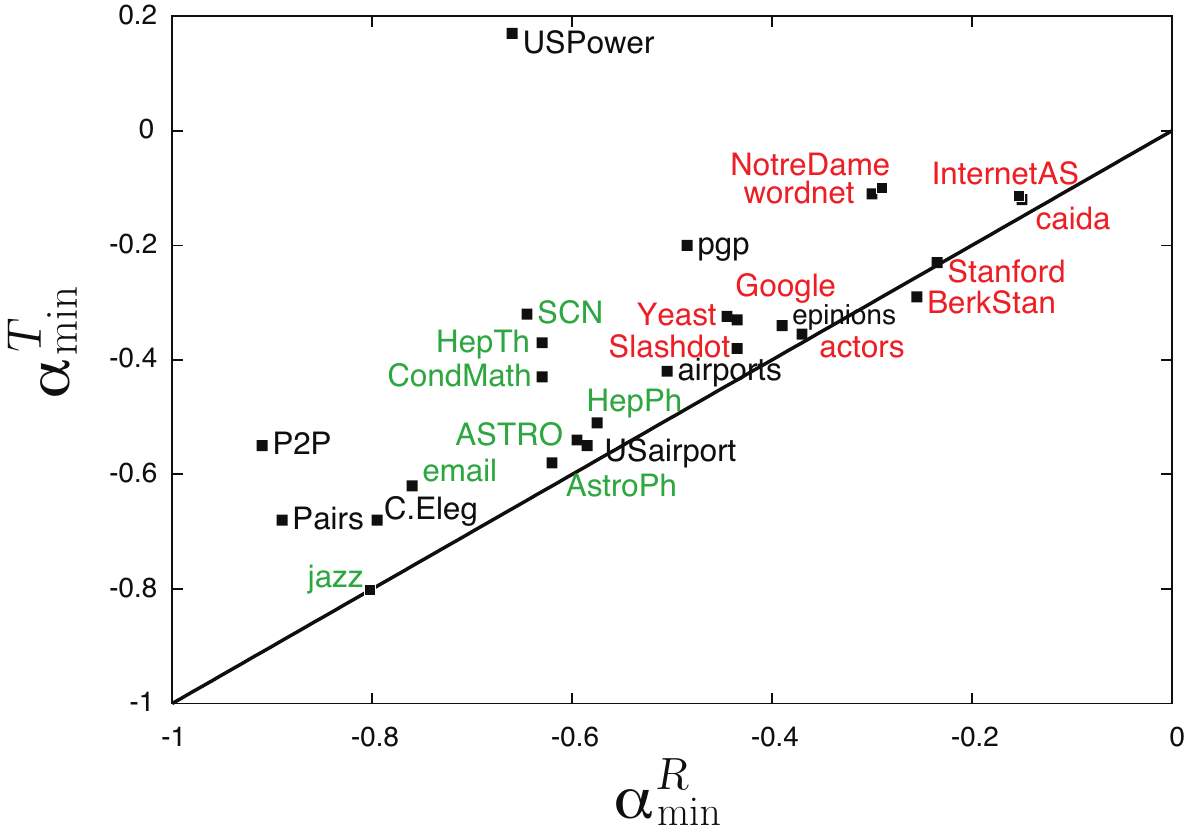}
   \label{fig:mrt_vs_mfpt}
  }  
 \caption{(color online) (a) The graph mean first passage time $T$,
   rescaled by the number of nodes $N$, plotted as a function of
   $\alpha$ for SCN (dotted green line), InternetAS (dashed red line)
   and an uncorrelated scale-free graph with $N=10^4$ and $\gamma =
   2.5$ (solid blue line). (b) The value $\alpha^T_{\rm min}$ and the
   corresponding minimum value of global mean first passage time
   $T_{\rm min}/N$ for all the networks in the data set. (c) There is
   a strong positive correlation between the two values of the bias
   $\alpha$ which minimize respectively MRT and MFPT. The solid line
   corresponds to $\alpha_{\rm min}^{R}=\alpha_{\rm min}^{T}$. The
   value of the Pearson's linear correlation coefficient is $r=0.87$
   (the US power-grid network is excluded). }
\end{figure*}

Differentiating Eq.~(\ref{eq:mrt_meanfield_integral}) with respect to
$\alpha$ we get again the value $\alpha^{R}_{\rm min} = -1$, while the
second derivative is always positive, as expected.  It is worth
noticing that the result $\alpha^{R}_{\rm min}=-1$ is independent from
the value of the scaling exponent $\gamma$ of the degree distribution
and from the maximum degree in the network, $k_{max}$.
%
%
%

The quantity $R$ is an average over all graph nodes.  However,
Eq.~(\ref{eq:mrt_k_meanfield}) allows also to compute the value
$\alpha^{r}_{\rm min}(k)$ that minimizes the return time $r(k)$ for nodes
having a certain degree $k$.
In the case of \Erdos--R\'enyi graphs a large number of nodes have the
same degree because the degree distribution is picked around $\langle
k \rangle$ and, as a result, the values of return times are very
similar for most of the nodes.  For real-world networks, instead, the
degree distribution is often heterogeneous and the the return time
sensibly depends on the degree of the starting node.
Differentiating Eq.~(\ref{eq:mrt_k_meanfield}) with respect to
$\alpha$ we get:
\begin{equation}
0=  \frac{ d } { d \alpha} r(k)= C'_{\alpha} k^{-\alpha-1} -
  C_{\alpha} k^{-\alpha -1} \log(k)
  \label{eq:tau_derivate_k}
\end{equation}
being $C_{\alpha} = N  \avg{k^{\alpha+1}} $.

Replacing the average over nodes $\avg{\ldots}$ with the integral over
degree classes, and considering networks with power-law degree
distributions $P(k) \sim k^{- \gamma}$ and with minimum and maximum
degree $k_{m}$ and $k_{M}$ we get
\begin{equation}
  C_{\alpha} \sim N \int_{k_{m}}^{k_{M}} k^{-\gamma} k ^{\alpha+1} \, dk
\label{eq:integrl_ca}
\end{equation}
Integrating Eq.~(\ref{eq:integrl_ca}) and plugging in
Eq.~(\ref{eq:tau_derivate_k}) we obtain:
\begin{equation}
\begin{split}
  & \Big[ \Big( k_{M}^{\beta} \ln k_{M} - k_{m}^{\beta}
    \ln k_{m} -  (k_{M}^{\beta} - k_{m}^{\beta}) \ln k
    \Big)\beta +\\
   &  + k_{M}^{\beta} + k_{m}^{\beta} \Big]
  k^{-\alpha -1}=0
\end{split}
\label{eq:a_min_of_k_cutoff}
\end{equation}
where $\beta = -\gamma + \alpha + 2 $.  The return time $r(k)$ for
nodes of a given degree class $k$ takes its minimum at the value of
$\alpha$ which satisfies the previous equation.  Excluding the
indeterminate case $\beta=0$, Eq.~(\ref{eq:a_min_of_k_cutoff}) has
only one solution for each value of $k$.

In the three top panels of Fig.~\ref{fig:figure2} we report the return
time $r(k)$ as a function of $\alpha$ for different degree classes
(solid lines), compared with the average return time $R$ of the same
graph (black dotted lines). The three panels correspond, respectively,
to (a) a configuration model scale-free graph with $\gamma=3$, (b)
InternetAS and (c) SCN. These plots show that a wrong choice of the
biased parameter can result in a large increase of the return
time. For instance in Fig.~\ref{fig:figure2} (c) the minimum return
time $r_{\rm min}(17)$ for the degree class $k=17$ occurs for $\alpha=0.5$
and is about four times smaller than the return time $r(17)$ obtained
at $\alpha=-1$ (refer to the vertical dashed lines for guidance).  

In the three bottom panels of Fig.~\ref{fig:figure2} we report, as a
function of $k$, the value $\alpha_{\rm min}^r(k)$ which optimized the
MRT for nodes having degree $k$. The black crosses are the numerical
results, while the solid blue line is the prediction in mean-field
obtained from the zeros of Eq.~(\ref{eq:a_min_of_k_cutoff}).  We
notice an excellent agreement between the numerical results and the
mean-field solution in the case of the uncorrelated scale-free graph
(panel (d)), while for real-world networks (panel (e) and (f)) we
observe considerable deviations from the analytical prediction,
evidently due to the presence of degree-degree correlations.  From the
point of view of network exploration, Eq.~(\ref{eq:a_min_of_k_cutoff})
turns out to be useful when an agent is sent through the network in
order to collect information and then has to come back to its starting
point~\cite{Prignano2012}. In fact, this equation gives insight about
how to fine-tune the bias parameter in order to increase or decrease
the time required (on average) by the agents to come back to the
starting nodes with the collected information.  It is worth noticing
that small changes in $\alpha$ can produce large variations in the
return times, thus highlighting the importance of a proper tuning of
the motion bias.
%

\section{Mean First Passage Time}
\label{section:mfpt}

In this Section we focus on the mean first passage time, showing that
the interplay between degree correlations and the dynamics of biased
random walks produces qualitatively similar results to those found for
the mean return time.

We denote as $t_{ij}$ the expected time needed for a random walker to
reach node $j$ for the first time when starting from node $i$.  If the
transition matrix $\Pi$ of the walker is primitive, it is possible to
determine $t_{ij}$ by using the fundamental matrix of the Markov chain
associated to the random walk~\cite{grinstead}. The fundamental matrix
$Z$ is defined as:
\begin{equation}
  Z = (I - \Pi^\top + W)^{-1}
  \label{eq:fundmatrix}
\end{equation}
where each row of $W$ is equal to the stationary probability
distribution $\bm{p^*}$ and $I$ is the identity matrix. The mean
first passage time $t_{ij}$ is then equal to:
\begin{equation}
  t_{ij} = \frac{z_{jj}-z_{ij}}{p^*_j}
  \label{eq:FPT}
\end{equation}
where $z_{jj}$ and $z_{ij}$ are the entries of the fundamental matrix $Z$.
Notice that in general $t_{ij} \neq t_{ji}$.
We define the {\it graph mean first passage time} $T$ as the average 
of the first passage time over all possible node pairs:
\begin{equation}
   T = \frac{1}{N(N-1)} \sum_{i,j} t_{ij}
\end{equation}
Notice that in general the calculation of the fundamental matrix in
Eq. (\ref{eq:fundmatrix}) is computationally intensive, since it
requires the inversion of a $N\times N$ matrix, and is practically
unfeasible for large networks. For this reason we used the fundamental
matrix $Z$ only to compute the mean first passage time for relatively
small networks ($N \lesssim 10^4$), while we resorted to agent-based
simulation for larger networks (see Appendix for a description of the
employed agent-based algorithm).

As found for the global mean return time $R$, also $T$ is a convex
function of the bias parameter $\alpha$ with a single minimum at
$\alpha^{T}_{\rm min}$.  This is illustrated in
Fig.~\ref{fig:mfpt_of_assort_disassor}.  Again, the position of the
minimum is at $\alpha=-1$ only for uncorrelated networks (see
Table~\ref{tab:list_net}).  We also notice that for disassortative
real-world networks $-0.5\ < \alpha^{T}_{\rm min}< 0$ as already found
in the case of the mean return time. Conversely, some assortative
networks can have a value $\alpha^{T}_{\rm min}$ which is not in the
range $[-1,-0.5]$. It is worth noticing that the minimum value $T_{\rm
  min}$ in real-world networks is significantly smaller than the MFPT
for unbiased ($\alpha=0$) random walks, or for the case of
uncorrelated networks ($\alpha=-1$).
In Fig.~\ref{fig:mfpt_rescaled} we plot, for all the networks in the
data set, the minimum value of the graph first passage time $T_{\rm
  min}$ rescaled by the number of nodes $N$. Despite there is no clear
separation at $\alpha=-0.5$ between assortative and disassortative
networks, as observed for the MRT, the behavior is similar to that
shown in Fig.~\ref{fig:mrt_rescaled}: the farther $\alpha^{T}_{\rm
  min}$ gets from $-1$, the more $T_{\rm min} / N$ deviates from $1$.

\begin{figure*}[t]
\centering
  \subfigure[]{
  \includegraphics[width=2.23in]{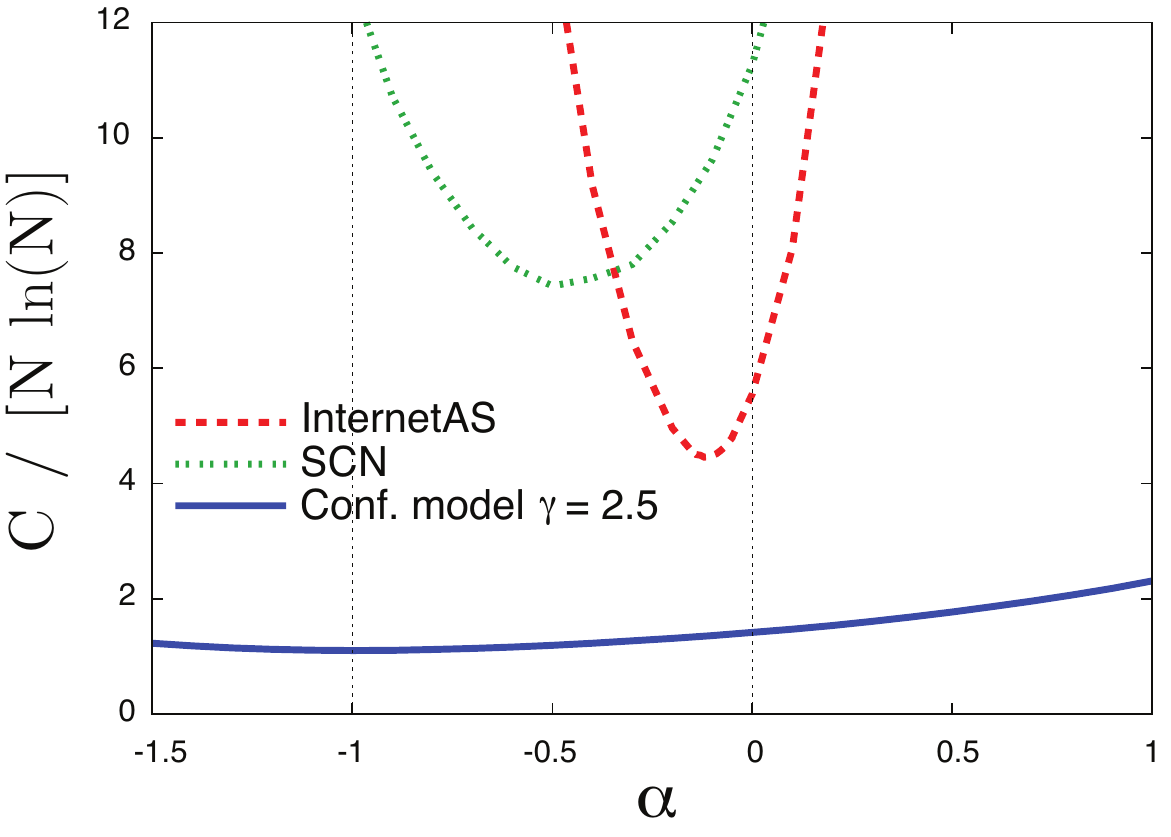}
  \label{fig:mct_of_assort_disassor}
  }
  \subfigure[]{
  \includegraphics[width=2.19in]{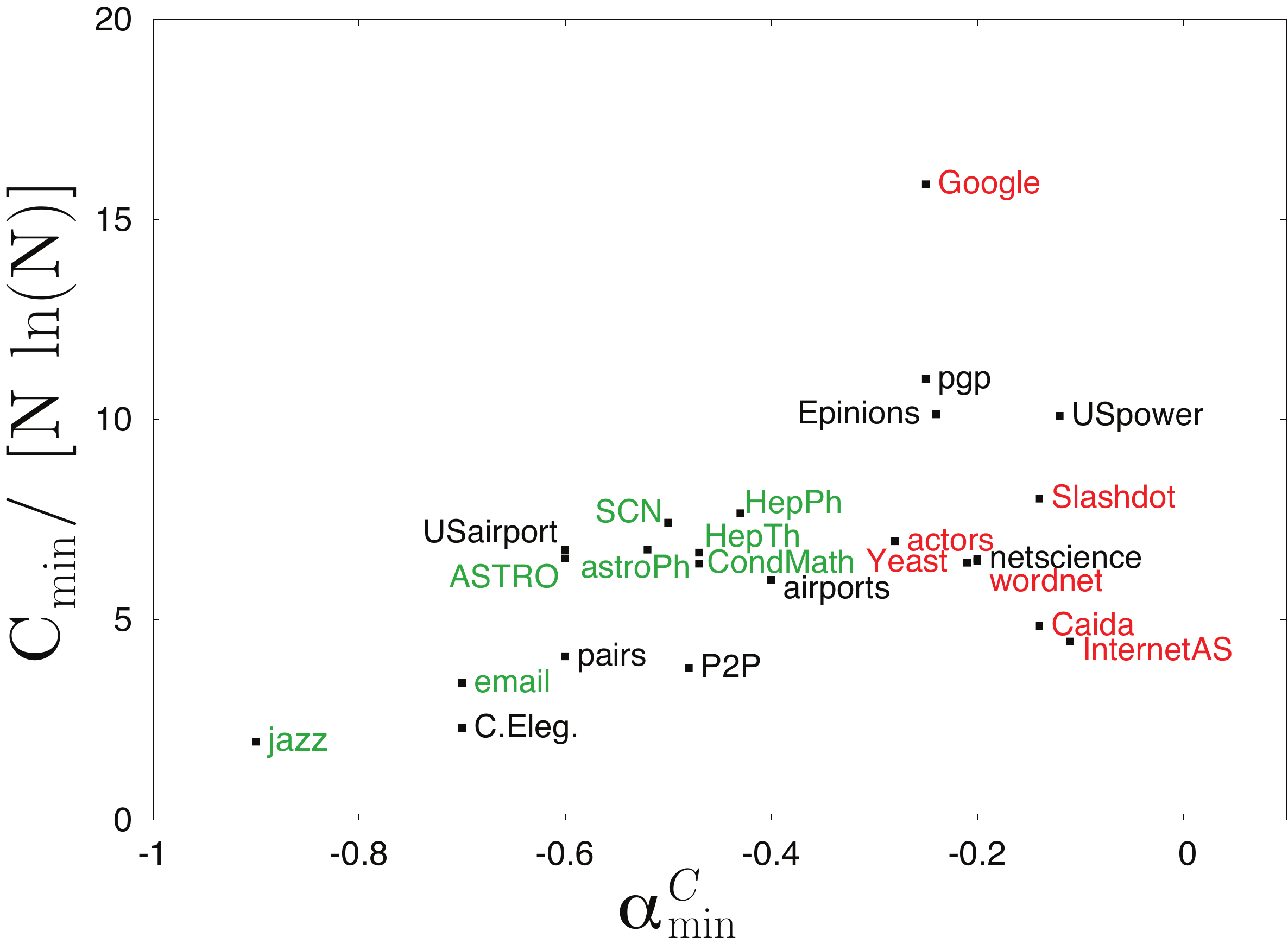}
    \label{fig:mct_rescaled}
  }
  \subfigure[]{
  \includegraphics[width=2.29in]{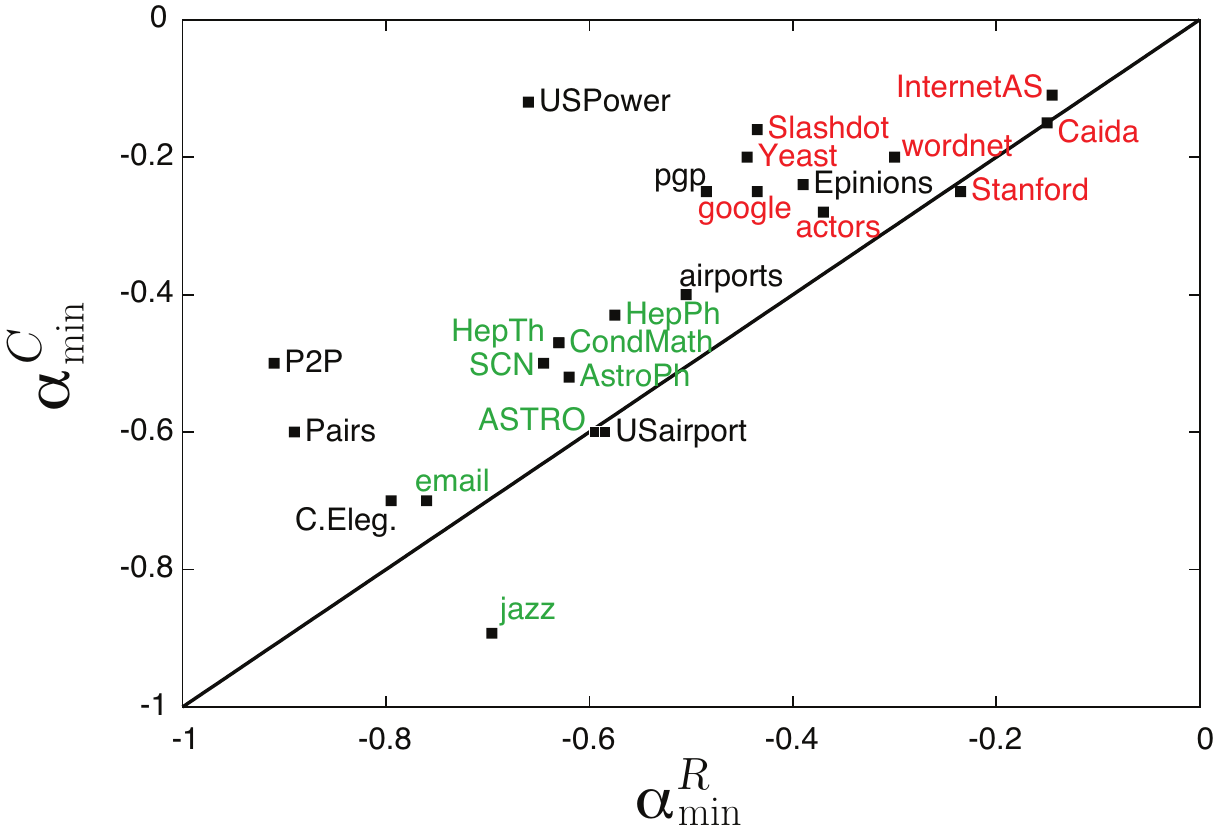}
  \label{fig:MRT_VS_MCT}
  }
 \caption{(color online) (a) The graph mean coverage time $C$,
   rescaled by the lower bound $N \ln (N)$, plotted as a function of
   $\alpha$ for the same networks as in
   Fig.~\ref{fig:mfpt_of_assort_disassor}.  (b) The minimum value
   $\alpha^C_{\rm min}$ and the corresponding coverage time $C_{\rm
     min}$ for all the networks in the data set. (c) There is a
   positive correlation between the two values of the bias $\alpha$
   which minimize respectively MRT and MCT. The solid line corresponds
   to $\alpha_{\rm min}^{R}=\alpha_{\rm min}^{C}$. The value of the
   Pearson's linear correlation coefficient is $r=0.77$ (the US
   power-grid network is excluded).}
\end{figure*}

A comparison between the values of $\alpha_{\rm min}$ for MFPT and MRT
is shown in Fig.~\ref{fig:mrt_vs_mfpt}. Excluding the network of the
US power grid (indicated in the figure as USPower), the value of the
Pearson's linear correlation coefficient between $\alpha^R_{\rm min}$
and $\alpha^T_{\rm min}$ is $r= 0.87$. Despite the two values of
$\alpha_{\rm min}$ are not equal for all networks, the strong positive
correlation we observe is quite remarkable. We notice that the US
power grid is the only spatially embedded network in the data set, so
that its exceptional values of $\alpha_{\rm min}$ can be due to
spatial constraints, which are absent in the other networks
studied. The existence of a relatively strong positive correlation
between $\alpha^R_{\rm min}$ and $\alpha^T_{\rm min}$ could have
interesting practical applications. In fact, in order to obtain a walk
having a small graph MFPT on a large network, it is possible to use
$\alpha^R_{\rm min}$ as an approximation of $\alpha^T_{\rm min}$, so
that one can obtain a quasi-optimal biased random walk with respect to
MFPT without the need to invert the fundamental matrix of the graph,
which is practically impossible for large networks.

\section{Mean coverage time}
\label{section:mct}

The last characteristic time under investigation is the mean coverage
time (MCT) $c_i$, defined as the expected number of time steps
required for the walker to visit all the nodes of the graph at least
once when starting from node $i$.  We also study the {\it graph mean
  coverage time} $C$, defined as an average of $c_i$ over all the
graph nodes:
\begin{equation}
\centering
C = \frac{1}{N} \sum_i^N c_i
  \label{eq:MCT}
\end{equation}
We have computed the graph mean coverage time $C$ for all but two
networks in the data set by means of an agent-based simulation and by
averaging over many realizations of the walk as described in the
Appendix.
The asymptotic lower bound on the coverage time for the unbiased
($\alpha=0$) random walk on a generic graph is given by
\cite{Feige1995}:
\begin{equation}
\centering
c_i \ge (1+O(1)) \, N \, \ln (N)   \,\,\,\,\,\,\,   (\alpha=0)
\label{eq:boundMCT}
\end{equation}
where the equality is satisfied for the complete graph of $N$ nodes, 
i.e. the graph in which there is a link between every pair of nodes. 
The inequality (\ref{eq:boundMCT}) implies the following lower bound for the global
mean coverage time:
\begin{equation}
\centering
C  \ge  (1+o(1)) \, N \, \ln (N) \,\,\,\,\,\,\,   (\alpha=0)
\label{eq:lower_bound}
\end{equation}
We therefore normalize the obtained values of $C$ by the quantity $ N
\, \ln (N)$. 

In Fig.~\ref{fig:mct_of_assort_disassor} we report such normalized
quantity as a function of the bias parameter for a configuration model
scale-free network, SCN and InternetAS. The mean coverage time is a
convex function of $\alpha$ with a single minimum at $\alpha^C_{\rm
  min}$.  As for MRT and MFPT we notice that the minimum of the global
mean coverage time for the uncorrelated scale-free graph occurs at
$\alpha^C_{\rm min}=-1$, and that the minimum value $C_{\rm min}$ is
very close to the lower bound given by Eq.~(\ref{eq:lower_bound}).
Real-world networks have instead values of $C_{\rm min}$ significantly
higher than the lower bound.

We notice that the MCT is more sensitive to $\alpha$ than MRT and MFPT
(the typical concavity of MCT in Fig.~\ref{fig:mct_of_assort_disassor}
is narrower than the ones observed for MRT and MFPT, respectively in
Fig.~\ref{fig:mrt_of_assort_disassor} and in
Fig.~\ref{fig:mfpt_of_assort_disassor}). For instance, in SCN the
minimum mean coverage time, $C_{\rm min} \simeq 7 N ln(N)$, is about $1.7$
times smaller than the mean coverage time obtained for $\alpha=0.0$ or
for $\alpha=-1$ on the same graph, which is $C_{(\alpha=0)} \simeq
C_{(\alpha=-1)} \simeq 12 N ln(N)$. Instead, disassortative networks
like InternetAS have a minimum value of the coverage time that is
similar to that for the unbiased case, while extremely different from
the value at $\alpha=-1$.

In Fig.~\ref{fig:mct_rescaled} we report the values of
$\alpha^{C}_{\rm min}$ and $C_{\rm min}$ for all the networks in the
considered data set. The results are qualitatively similar to those
reported in Fig.~\ref{fig:mrt_rescaled} and
Fig.~\ref{fig:mfpt_rescaled}.
Although for a given network the minimum of $C$ occurs at
$\alpha^{C}_{\rm min} \ne \alpha^{R}_{\rm min}$, it is evident from
Fig.~\ref{fig:MRT_VS_MCT} that the two values are positively
correlated (the Pearson's linear correlation coefficient is $r=0.77$).

We have investigated the differences between the
  optimal values of $\alpha$ for the three characteristic times
  comparing $\alpha_{\rm min}^R$, $\alpha_{\rm min}^T$ and
  $\alpha_{\rm min}^C$ for a set of synthetic networks generated
  through the swapping procedure (the starting network in this case is
  a configuration model with $\gamma=2$, $N=1000$ and $\langle k
  \rangle=14.6$).  The results (figure not shown) suggest that
  synthetic assortative networks have equal optimal bias values
  ($\alpha_{\rm min}^R = \alpha_{\rm min}^T = \alpha_{\rm min}^C$), so
  that the deviations from the bisector in Figures
  \ref{fig:mrt_vs_mfpt} and \ref{fig:MRT_VS_MCT} might be due only to
  fluctuations in the pattern of degree correlations of real-world
  networks.  Instead, in the case of synthetic disassortative networks
  we observe deviations from the bisector of the same order of those
  observed in real-world networks.

\section{Discussion}
\label{section:discussion}

In this section we discuss in detail some of the
  results reported in the paper, we provide a mechanistic explanation
  of the variations of $\alpha_{\rm min}$ observed in real-world
  networks and we outline possible applications to practical
  problems.

\textit{Deviations from $\alpha_{\rm min}=-1$. ---}
  The results reported in Fig.~\ref{fig:mrt_rescaled} confirmed that
  the value of $\alpha$ which minimizes the MRT in real-world networks
  sensibly deviates from the value $\alpha_{\rm min}=-1$ predicted for
  uncorrelated graph, and that this deviation seems to depend on the
  sign and magnitude of degree-degree correlations.  We notice that,
  in the absence of degree correlations, the stationary distribution
  of walkers is given by Eq.~(\ref{eq:pstar_mean_field}), which for
  $\alpha=-1$ corresponds to a uniform distribution of walkers across
  the nodes of the network, i.e. $p^*_i=1/N$. Consequently, the
  minimum value of MRT for uncorrelated graphs is obtained for a
  uniform distribution of walkers and is equal to $R_{\rm min}=N$ (see
  Eq.~(\ref{eq:mrt_meanfield})). We argue that the minimum of MRT in a
  generic network is always obtained for a value of $\alpha$ which
  induces the distribution of walker that is the closest possible to a
  uniform one.

We start by noticing that, according to
  Eq.(\ref{eq:net_meanRT}), $R$ is the harmonic mean of the stationary
  distribution $\bm{p^*}$. By using Jensen's
  inequality~\cite{Jensen1906}, it is possible to prove that any
  stationary distribution $\bm{p^{*}}$ which is not uniform produces a
  value of the mean return time which is larger than (or at most equal
  to) that obtained from
  a uniform $\bm{p^*}$ (which is equal to $N$):
  \begin{equation}
	  R = \frac{1}{N} \sum_i \phi (p^*_i)  \geq  \phi \left(  \frac{
      \sum_i p^*_i }{N} \right) = N
  \end{equation}
where $\phi(x)=1/x$. We observe that if a graph is not uncorrelated,
and especially if the graph has assortative degree correlations, then
the stationary distribution of the biased random walk obtained for
$\alpha=-1$ is generally far from being uniform, while the stationary
distribution corresponding to $\alpha=\alpha_{\rm min}$ is usually very
close to a uniform one.  And in fact, Fig.~\ref{fig:mrt_rescaled}
confirms that for assortative networks the value of $R_{\rm min}$ is very
close to $N$, despite larger deviations are observed for
disassortative networks. Thus we assume that, for a given network, the
discrepancy between the observed value of $\alpha_{\rm min}$ and the
prediction $\alpha=-1$ for uncorrelated networks is indeed due to the
necessity to obtain a stationary distribution as close as possible to
a uniform one.

\begin{figure}[!t]
	\includegraphics[width=3in]{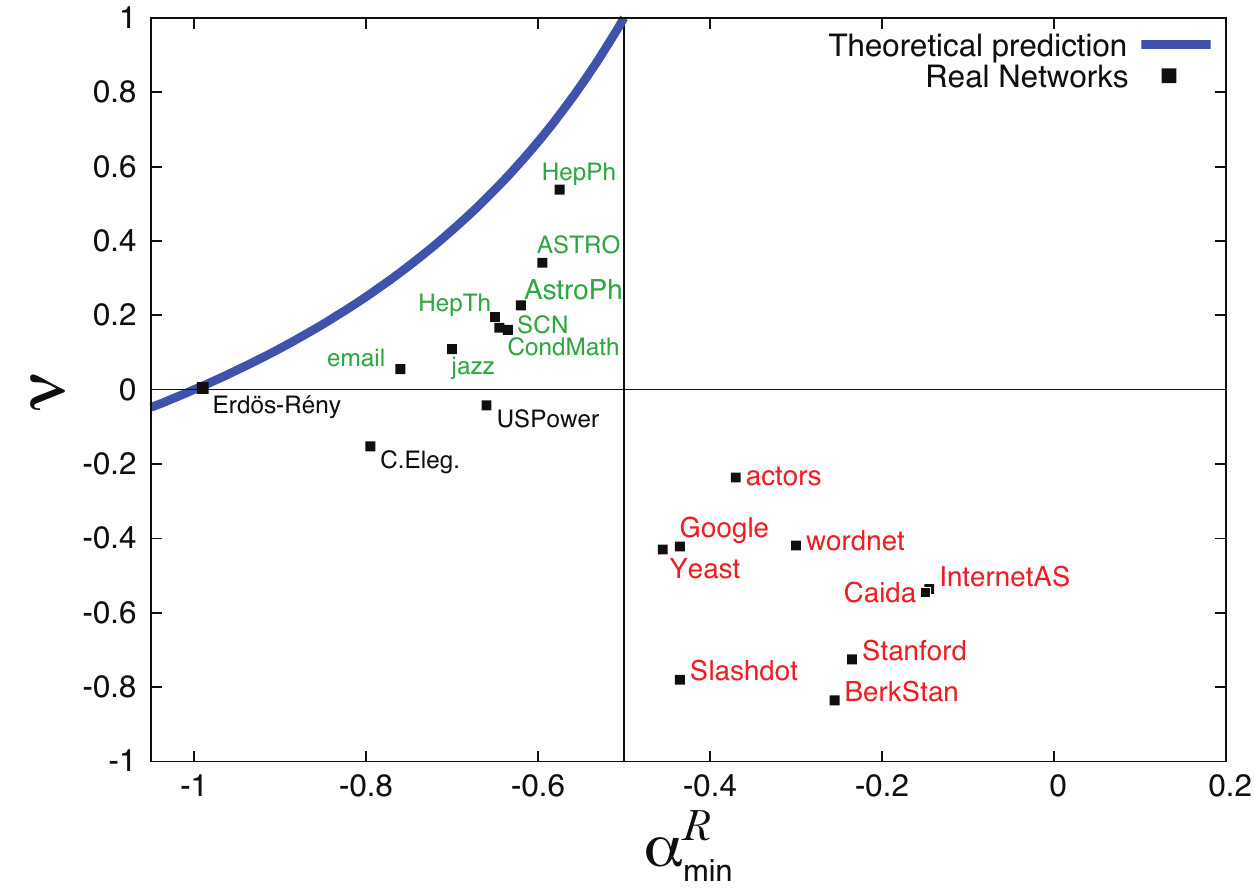}
	\caption{(color online) The theoretical prediction (solid blue
    curve) of $\alpha_{\rm min}$ given by Eq. (\ref{eq:a_nu_relation})
    works well for assortative networks ($\nu=1$ and $\alpha_{\rm
      min}=-0.5$) and uncorrelated networks ($\nu=0$ and $\alpha_{\rm
      min}=-1$).  The theoretical curve falls outside the plot for
    negative values of $\nu$ and does not approximate well the optimal
    bias value for disassortative networks.}
	\label{fig:exponent_pstar_stat}
\end{figure}

If this hypothesis is correct, it should be possible
  to determine the value of $\alpha_{\rm min}$ by imposing that the
  resulting stationary occupation probability distribution is as close
  as possible to $p_i=1/N$. Let us consider the case of assortative
  networks, and assume that the expected degree of the first
  neighbors of a node having degree $k$ is a power-law, i.e.:
  \begin{equation}
    k_{nn}(k)=D k^{\nu}, \quad \nu > 0
    \label{eq:knn_theo}
  \end{equation}
  where $D$ is a normalization constant. Let us also make the
  assumption that the fluctuations in the degree of the neighbors of
  a node with degree $k$ are negligible, so that if $j$ is a first
  neighbor of node $i$ we can write:
  \begin{equation*}
    k_j \simeq k_{nn} (k_i) = D k_i ^ \nu
  \end{equation*}
  By plugging Eq.~(\ref{eq:knn_theo}) in Eq.~(\ref{jesus_prob}) we get:
\begin{equation}
	c_i \sim k_i^{\alpha \nu} \sum_j a_{ij}=k_i^{\alpha\nu + 1} \, , \;\;\;\;\ p^*_i \sim
  \frac{k_i^{\alpha \nu + \alpha +1}} { \sum_\ell k_\ell^{\alpha \nu +
      \alpha +1}}
	\label{eq:pstar_average}
\end{equation}
Imposing that $\bm{p^*}$ is a uniform distribution, i.e. that $p^*_i =
p^*_j, \> \forall \> i,j=1,\ldots,N$, we obtain that $\alpha$ should
satisfy the equation:
\begin{equation}
	\alpha \nu + \alpha +1 = 0, \quad \nu >0.
	\label{eq:a_nu_relation}
\end{equation}
In Fig.~\ref{fig:exponent_pstar_stat} we show the curve $\alpha\nu +
\alpha + 1 = 0$ (solid blue line) together with the values
$(\alpha_{\rm min},\nu)$ corresponding to real-world networks. Notice
that for uncorrelated networks, i.e. when $\nu=0$, we obtain the
analytical prediction $\alpha_{\rm min}=-1$, while for maximally
assortative networks, i.e. for $\nu=1$, we get $\alpha_{\rm
  min}=-0.5$. Also, the values of $(\alpha_{\rm min}, \nu)$ for
real-world assortative networks are close ---but admittedly not
identical--- to the prediction of Eq.~(\ref{eq:a_nu_relation}). The
observed discrepancies between theory and observations are due to the
fact that, despite in real-world networks we usually have
$k_{nn}(k)\sim k^{\nu}$, the fluctuations in the degree of the nearest
neighbors of a node with degree $k$ are not negligible. Therefore, if
$j$ is a neighbor of $i$ then $k_j \neq k_{nn}(k_i) \sim k_i^{\nu}$,
and consequently the second of the two assumptions used in the
derivation of Eq.~(\ref{eq:pstar_average}) and
Eq.~(\ref{eq:a_nu_relation}) does not hold.

 The case of disassortative networks is a bit more
  cumbersome, due to the constraints introduced by a discrete degree
  sequence (namely, a node cannot have a degree smaller than $1$ or
  larger than $k_{max}$). In particular, it is possible to define:
  \begin{equation}
    k_{nn}(k) \sim  k^{\nu} k_{max}, \quad \nu < 0
  \end{equation}
obtaining an equation similar to
Eq.~(\ref{eq:a_nu_relation}). Unfortunately, such equation has a
discontinuity at $\nu=-1$ and does not match the values of
$(\alpha_{\rm min}, \nu)$ observed in real-world disassortative
networks.

\begin{figure}[!t]
	\includegraphics[width=3in]{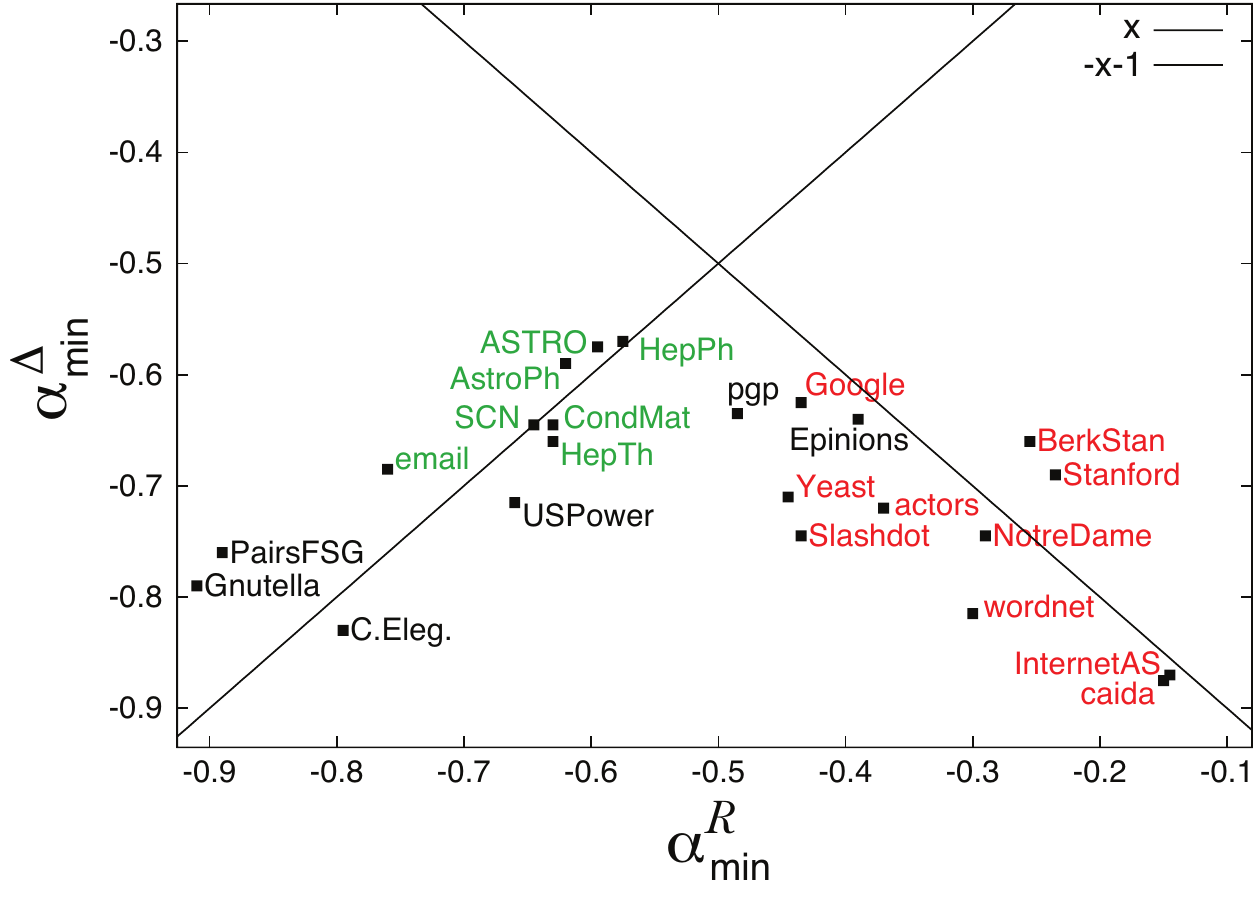}
	\caption{(color online) The value $\alpha_{\rm min}^{\Delta}$ of the
    motion bias whose corresponding stationary distribution is the
    closest one to a uniform distribution is correlated with
    $\alpha_{\rm min}^{R}$. In particular, in assortative networks
    (green labels) the two values are positively correlated
    ($\alpha_{\rm min}^{R} \simeq \alpha_{\rm min}^{\Delta}$), while
    in disassortative networks we have $\alpha^{\Delta}_{\rm min}
    \simeq -1 - \alpha_{\rm min}^{R}$.}
	\label{fig:mrt_vs_variance}
\end{figure}

 In the absence of an analytical argument for
  disassortative networks, we computed numerically the value of
  $\alpha$ which minimizes the variance $\Delta = \avg {p^2_i} - \avg
  {p_i}^2 $ of the stationary occupation probability distribution. In
  fact, $\Delta$ provides a rough estimation of how far the
  distribution is from a uniform one (for which $\Delta=0$). The
  results are reported in Fig.~\ref{fig:mrt_vs_variance}, in which we
  show, for each network in the considered data set, the value of
  $\alpha_{\rm min}^{R}$ which minimizes the MRT and the value
  $\alpha_{\rm min}^{\Delta}$ which minimizes the variance of
  $\bm{p^*}$. Notice that for assortative networks we have a strong
  positive correlation between $\alpha_{\rm min}^{R}$ and
  $\alpha_{\rm min}^{\Delta}$, with $\alpha_{\rm min}^{\Delta} \simeq
  \alpha_{\rm min}^{R}$. Conversely, for disassortative networks the two
  values are negatively correlated, and seem to be connected by the
  relation $\alpha_{\rm min}^\Delta \simeq -1 - \alpha_{\rm
    min}^R$. These results confirm that there is indeed an intimate
  relation between the variance of the stationary state distribution
  of walkers obtained for a given value $\alpha$ of the motion bias
  and the corresponding MRT, and suggest that the optimization of the
  mean return time is obtained for a value of $\alpha$ which
  guarantees a stationary distribution as close as possible to a
  uniform one.

\textit{Applications. ---} The results shown in this
  paper can have interesting applications in several different
  contexts, from the control of diffusion processes to the successful
  advertisement of products and services on online social networks. A
  typical example is that of congestion control in communication and
  transport systems, such as the Internet, the WWW, P2P networks and
  road networks. In these systems traffic is usually modeled through
  simple packet generation and routing algorithms. At each time step,
  a certain number $R$ of new packets is introduced in the system, and
  each packet is assigned a source and a destination node. The nodes
  of the network route packets according to a certain policy (which
  might be a biased random walk), and have a fixed delivery capacity
  $D$. When a packet arrives at its destination node, it is removed
  from the system.  An interesting quantity that characterizes the
  emergence of congestion is the critical packet generation rate
  $R_c$, defined as the number of new packets above which the number
  of packets removed from the system per unit time is smaller than the
  number of new packets per unit time introduced in the network.
  Under the condition $R>R_c$ the number of packets flowing in the
  network keeps increasing with time, leading to congestion. In
  Ref.~\cite{Fronczak2012} it has been shown that, for a routing
  strategy based on biased random walk, $R_c$ and the graph mean first
  passage time $T$ are related by the equations:
\begin{equation}
R_c (i) =  \frac{D} { p^*_i T} 
\label{eq:packetsrate1}
\end{equation}
\begin{equation}
R_c = \min_i \left\{ R_c(i) \right\}
\label{eq:packetsrate2}
\end{equation}
where $D$ is the delivering capacity.  Eq. (\ref{eq:packetsrate1})
shows that the value of $R_c$ depends on $\alpha$ through both $T$ and
$p^*_i$.  This implies that in order to maximize $R_c$ one has to
minimize the product $ \displaystyle \max_i\{p^*_i\} \, T$. By
noticing that, in assortative networks, for $\alpha=\alpha_{\rm
  min}^{R}$ we obtain a $\bm{p^*}$ which is the closest possible to a
uniform distribution (according to the results shown in
Fig.~\ref{fig:mrt_vs_variance}) and, at the same time, we get an
almost minimal value of $T$ (due to the strong positive correlation
between $\alpha_{\rm min}^R$ and $\alpha_{\rm min}^T$), we can
conclude that a good approximation for the value of the bias which
maximizes $R_c$ can be obtained by setting $\alpha=\alpha_{\rm
  min}^{\Delta}\simeq \alpha_{\rm min}^{R}\simeq \alpha_{\rm
  min}^{T}$. Unfortunately, this reasoning does not work in
disassortative networks, for which the value of $\alpha$ that
minimizes $\max_i\{p^*_i\}$ does not coincide with the value of alpha which
minimizes $T$. In this case, the optimal critical packet rate depends
on the trade-off between the homogeneity of the stationary occupation
probability distribution and the corresponding value of $T$.
This is an example of how the correlation between the optimal values
of $\alpha_{\rm min}$ and the sign and magnitude of degree-degree
correlations can be used to avoid congestion and improve transport
performance on a given network.

The results of this paper might also find application
  in the field of optimal network crawling, i.e. the exploration of
  the structure of a graph by means of agents performing random walks
  over it. Examples include the sampling of online social networks
  (e.g., Facebook and co-purchasing networks) and online communication
  networks (e.g., the World Wide Web and Twitter).  In particular,
  exploring the network at the fastest possible speed corresponds to
  minimizing the MCT. As we have seen, this is achievable by using a
  degree-biased random walk with $\alpha=\alpha_{\rm min}^{R}$, since
  there is a pretty strong correlation between $\alpha_{\rm min}^{R}$
  and $\alpha_{\rm min}^{C}$. If the network is assortative, which is
  actually the case for the majority online social networks, the value
  of $\alpha_{\rm min}^{R}$ which optimizes the coverage time will lie
  in $[-1,0.5]$ and can be obtained using Eq.~(\ref{eq:a_nu_relation})
  where the correlation exponent $\nu$ can be measured from a
  relatively small sample of the graph of interest. If instead the
  network is disassortative, as usually happens for online
  communication networks, then the value of $\alpha$ should be chosen
  in the range $[-0.5,0]$ and a good hint is provided by the value $-1
  - \alpha_{\rm min}^{\Delta}$ (see
  Fig.~\ref{fig:mrt_vs_variance}). Such value can be computed taking
  into account a small representative sample of the degree sequence of
  the graph. In both cases, an appropriate tuning of the bias
  parameter $\alpha$ will outperform the standard unbiased random
  walk.

Another interesting application of the relationship
  between assortativity and optimal graph traversal could be that of
  information retrieval. In a recent work~\cite{kivimaki2013} it has
  been shown that the biased random walk on the directed network of
  Wikipedia pages can be used to implement an algorithm able to
  retrieve professional skills from an arbitrary text (e.g., a
  curriculum vitae). The authors have shown that the performance of
  the system can be optimized by means of an appropriate tuning of the
  motion bias $\alpha$. The results reported in Table 2 of
  Ref.~\cite{kivimaki2013} show that the best performance of the
  retrieval system are achieved for $\alpha$ between $-1$ and $0$ and
  in particular for $\alpha\simeq -0.4$ which is a reasonable optimal
  value of the motion bias considering that the undirected version of
  the network of Wikipedia pages is known to be
  disassortative. Therefore, the generalization of the present study
  to the case of directed networks could provide theoretical insights
  and guidelines for the optimal choice of the bias parameter in
  skill retrieval system.

Finally, another possible application of these results
  concerns social-marketing campaigns. Today the advertising of
  products and services is more often conveyed through online social
  networking platforms. Customers are promised a reward if they
  promote a certain range of products to their on-line friends, and
  usually they get an equal reward for each friend that adopts the
  product/service. If we assume that the diffusion of the advertising
  can be regarded as a random motion, then promising equal rewards is
  not the best diffusion strategy, because customers will not have any
  reason to preferentially advertise the product to any of their
  neighbors in particular, and will therefore choose one of their
  friends at random, with equal probability. If we look to the
  advertisement as a walker which jumps from one customer to another,
  this strategy would correspond to an unbiased random
  walk($\alpha=0$). Our results about mean coverage time suggest
  instead that the diffusion speed (i.e. the number of advertised
  users per unit time) can be increased if the customer is rewarded
  proportionally to a biased transition probability, i.e. if the
  customer receives a reward proportional to the $\alpha$ power of the
  degree of the friend who has adopted the suggested product/service.
  The optimal bias parameter can be directly computed if the network
  topology is entirely known or, given that social networks are often
  assortative, it can be guessed using Eq.~(\ref{eq:a_nu_relation}).
  This also disproves the intuitive idea that the best strategy is to
  always advertise the highly connected users.

\section{Conclusions}
\label{section:conclusions}

Random walks are the simplest way to visit a network, and
degree-biased random walks, which make use of information about the
degree of destination nodes, are particularly suited to highlight the
presence of degree-degree correlations. In this paper we have focused
on the typical times of biased-random walks, namely on the expected
time that a walker needs to come back to its starting node (MRT), to
hit a given node (MFPT), or to visit all the nodes of the network
(MCT). We have studied how such characteristic times depend on the
value of the motion bias $\alpha$.  We have proved analytically that,
in the mean-field approximation, the value $\alpha_{\rm min}$ that
minimizes the characteristic times in uncorrelated networks is equal
to $-1$. This corresponds to a walk in which the probability to move
to a node is inversely proportional to its degree. As shown by
numerical simulations, the mean-field approximation works pretty well
for uncorrelated networks. However, real-world networks are
characterized by non-trivial degree-degree correlations and, as a
result, the characteristic times of degree-biased random walks on
real-world networks deviate from those obtained by using the
mean-field approach, as we have shown in the paper by studying a large
data set of medium to large real-world networks.

In particular, the value of $\alpha_{\rm min}$
  sensibly differs from $-1$, in a way that depends on the sign of the
  degree-degree correlations. We have found that optimal values of the
  bias parameter $\alpha$ lie between $-1$ and $0$ for a large number
  of real-world networks. In addition to this, we have shown that the
  minimum characteristic times occur preferentially for $\alpha$ in
  the range $[-1,-0.5]$ for assortative network, and for $\alpha$ in
  the range $[-0.5, 0]$ for disassortative ones. We have derived an
  approximate analytical relation between $\alpha_{\rm min}$ and the
  degree-correlation exponent $\nu$, which might be useful to refine
  the choice of the optimal bias for assortative networks, and we have
  shown numerically that the value of the mean return time obtained
  for a given value of $\alpha$ is related with the heterogeneity of
  the corresponding stationary probability distribution of the walk.

 By discussing several different possible applications
  of these results, we have stressed the fact that the minimization of
  characteristic times may be useful in many domains, from mitigation
  of network congestion to successful product advertisement in online
  social networks. In general, when only local information is
  available, degree-biased random walks can achieve better exploration
  performance than unbiased random walks, by appropriately tuning the
  bias parameter $\alpha$ according to the global structural
  properties of the graph at hand.

\section*{Appendix}

We describe here the algorithm we have used to generate graphs with
tunable degree-degree correlations, and the agent-based approach used
to estimate the mean coverage time and the mean first passage time in
large graphs.

\begin{figure}[ht]
	\centering
    \includegraphics[width=2.2in]{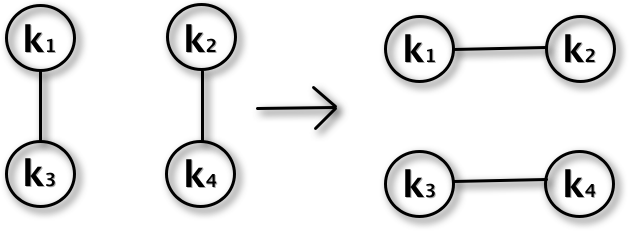}
    \caption{Assortative Swap}
        \label{fig:swappassort}
\end{figure}

\begin{figure}[ht]
	\centering
    \includegraphics[width=2.2in]{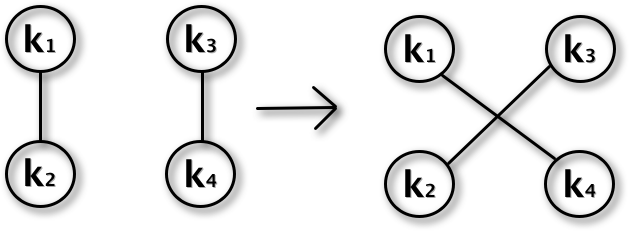}
        \caption{Disassortative Swap}
            \label{fig:swappdisassort}
\end{figure}

\bigskip
\noindent
\textbf{Swapping algorithm.}  In Fig.~\ref{fig:isteresi} and
Fig.~\ref{fig:isteresiMRT} we have reported the values of the
degree-correlations exponent $\nu$ and the motion bias which minimizes
the return time $\alpha_{\rm min}^{R}$ for a set of graphs with the
same degree sequence of a chosen starting graph and tunable
degree-degree correlations. An increasing amount of assortative or
disassortative correlations is introduced by repeatedly applying the
edge swapping procedure described in Ref.\cite{BrunetSokolov2005} to
an initially uncorrelated graph.  Each swap is performed as
follows. Two edges connecting four different nodes are randomly
selected and the nodes at the ends are ordered according to their
degree $k_1 \le k_2 \le k_3 \le k_4$.  The two edges are then removed.
Positive assortative correlations are introduced by connecting the two
nodes with the smaller degrees and the two nodes with the larger
degrees.  Instead, disassortative correlations are introduced by
connecting the node with the smallest degree with the node with the
largest degree and the two remaining nodes with intermediate degrees.
In order to preserve the degree sequence, all swaps that produce
parallel edges are not allowed.  Fig.~\ref{fig:swappassort} and
Fig.~\ref{fig:swappdisassort} illustrate the two types of swaps.

\bigskip

\noindent
\textbf{Agent-based simulation for MFPT and MCT.}	\\
The MFPT and MCT
are estimated by means of an agent based simulation.  In both cases we
simulated a walker which moves across the nodes of the network
according to the transition probability given in
Eq.~(\ref{eq:biased_trans_matrix}).
The simplest way to compute the characteristic times is to wait until
the walker, started at a randomly selected node, explores all the
nodes at least once.  At that point the value of $c_i$ is given by the
total number of time steps spent by the walker to visit all the nodes,
while $t_{ij}$ can be obtained by storing in memory the first passage
times to all other nodes during the simulation.  However, despite this
procedure is pretty simple to implement, it is not suitable to obtain
robust results in a reasonable amount of time. In fact, in order to
have an estimate of $c_i$ and $t_i$, we need to average over a
sufficiently large number of walks starting at node $i$, and the same
procedure should be repeated for all the starting nodes. However, the
heterogeneity of the degree distribution of real-world networks
induces heterogeneity in the number of visits on nodes with different
degree. Just to make and example, in the unbiased case $(\alpha=0)$
the walker visits a node with degree $1$ only once every $k_{max}$
visits on the node with the maximum degree.  As a result, most of the
computation time is wasted by repeated visits to highly connected
nodes. A value of $\alpha\neq 0$ can either accentuate or mitigate the
disproportion in the number of visits.  To overcome this problems, we
implemented a smarter strategy.  The key-point of our method is to
consider each hop as the starting point of a new walk and to store the
entire sequence of node labels in an array we call {\em Tape}.  As
soon as all nodes have been visited at least once, both $t_{ij}$ and
$c_i$ can be calculated (here $i$ is the node label at the beginning
of {\em Tape}). Then the first entry of {\em Tape} is removed, 
and the computation of the mean first passage and coverage time is
performed for the new node which now occupies 
the first entry of {\em Tape}.  If, after a removal
of the first entry, a node label is no longer contained in {\em Tape}
 new walker hops are simulated until all missing nodes are visited.

Here we describe separately the two algorithms for MCT and MFPT
despite the simulation could in principle be performed simultaneously.

\medskip
\noindent
\textbf{Algorithm for the Mean Coverage Time.} \\
We randomly select a
starting node and we simulate the walk according to the transition
probability of Eq. (\ref{eq:biased_trans_matrix}) for a given value of
$\alpha$.  We dynamically add the labels of the nodes visited at the
end of an array referred to as {\em Tape}.  An array {\em number-of-visits[i]} of
length $N$ keeps track of the number of visits on each node $i$. A
{\em counter} stores the number of unique nodes visited: when all nodes 
have been visited at least once the {\em counter} is equal to $N$. Finally a
variable {\em L} stores the number of hops between the node at the
first entry of {\em Tape} and the node at the end, i.e. the length
of {\em Tape} minus $1$.
The steps of the algorithm are reported in the following:

\begin{itemize}

\item[0)] Initialize all variables to zero and choose a node $i$ at random. 
Set {\em number-of-visits}$[i]$ and {\em counter} equal to $1$.

\item[1)] Jump to a successive node, say node $j$, and 
add the node label $j$ as new element at the end of 
{\em Tape} (push-back operation). Increase {\em L} and 
{\em number-of-visits}$[j]$ by $1$. If the new value of 
{\em number-of-visits}$[j]$ is equal to $1$ increase 
also {\em counter} by $1$.

\item[2)] If {\em counter} is equal to $N$ proceed 
to step 3) otherwise go to step 1).

\item[3)] The current vale of {\em L} is the estimate of 
the coverage time $c_i$ relative to the node in the first 
entry of {\em Tape} (let's say $i$). Store the value $c_i$ 
and the corresponding node label $i$.

\item[4)] Consider again the first entry $i$ of {\em Tape} and
  decrease {\em L} and {\em number-of-visits}$[i]$ by 1. If the 
  new value of {\em number-of-visits[i]} is equal to zero
  decrease also {\em counter} by $1$.

\item[5)] Remove the first entry $i$ of {\em Tape} and free
  the memory (pop-front operation). Then go to step 2).
  
\end{itemize}

\begin{figure} 
\centering
\includegraphics[width=3.2in]{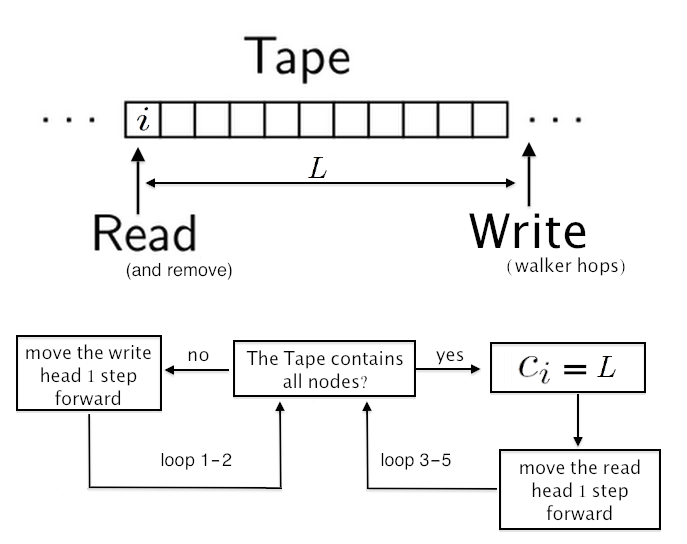}
\caption{The flowchart illustrates the core principle of the algorithm
  for the estimation of the mean coverage time. {\em Tape} is an array
  whose length changes dynamically. In steps 1-2 new node labels are
  written at the end of {\em Tape}, while in steps 3-5 nodes are
  removed from the beginning of {\em Tape}.}
\label{fig:flowchart}
\end{figure}

The simulation ends when the estimated values of $c_i$ are averaged
over at least 1000 realizations for each node $i$.  Consequently in
the unbiased case the value $c_j$ for a node $j$ with degree $k_j$
will be averaged over $1000 * k_j / k_{\rm min}$ realizations.  In
Fig. \ref{fig:flowchart} we illustrate the basic principle of the
algorithm.  The loop $1-2$ performs the walker motion and adds the
node labels in {\em Tape}.  When all nodes have been visited at
least once the algorithm enters in the $3-5$ loop where the estimates
of the coverage time are calculated and stored.  If the {\em
  number-of-visits[i]} for a certain node $i$ is equal to zero then
this node $i$ is no longer contained in {\em Tape} and the
algorithm goes back to the $1-2$ loop.

\medskip
\noindent
\textbf{Algorithm for the Mean First Passage Time.}\\
We notice that the estimation of the mean first passage time does not require the
computation of each entry of the matrix $Z$ but just the average of its
rows:
\begin{equation}
t_i = \frac{1}{(N-1)} \sum_{j\neq i} t_{ij}
\label{eq:mfpt_of_node}
\end{equation}
that is the average MFPT from node $i$ to all the other nodes.
We randomly select the starting node and we simulate the walk
according to the transition probability of
Eq. (\ref{eq:biased_trans_matrix}) for a given value of $\alpha$.
As before we add the labels of visited nodes at the end of {\em Tape}.
An array of dimension $N$ keeps track of the {\em number-of-visits}$[i]$ 
on each node $i$, and a {\em counter} stores the
number of unique nodes visited.  We use the variable {\em L} to keep
track of the total number of hops during the entire walk and in this
case this value will not be reduced when we pull off nodes from the
beginning of {\em Tape}.  Indeed we use a second variable
$L^{old}$ to store the number of nodes pulled off from {\em Tape}.
Moreover for each node $i$ we initialize an array {\em
  not-first-passage[i]} that store the times, i.e. the values of {\em
  L}, at which the walker visits a node already previously visited.
At later stages of the algorithm these values will be used to rapidly
compute the first passage time for a given walker path.  Finally, a 
variable {\em FPT} temporarily accumulates the sum of the values of
the first passages times $t_{ij}$ in order to calculate $t_i$ in
Eq. \ref{eq:mfpt_of_node}. Its role will be clear later.
The algorithm consists of the following steps:
\\
\begin{itemize}
\item[0)] Initialize all variables to zero and choose a 
node $i$ at random. Set {\em number-of-visits}$[i]$ 
and {\em counter} equal to $1$.

\item[1)] Jump to a successive node (let's say $j$), 
add the node label $j$ as new element at the end of 
{\em Tape} (push-back operation), and increase {\em L} by $1$.

\item[2)] If {\em number-of-visits}$[j]$ is equal to zero go to step
  3) otherwise go to step 4).

\item[3)] Add the value ({\em L} -{\em L$^{old}$}) to the variable
  {\em FPTs}. Increase the {\em counter} and {\em
    number-of-visits}$[j]$ by $1$. Then go to step 5).

\item[4)] Add the current value of {\em L} as new element at the end
  of the {\em not-first-passage}$[j]$ array and increase {\em
    number-of-visits}$[j]$ by $1$. Then go to step 5).

\item[5)] If {\em counter} is equal to $N$ go to step 6) otherwise go to step 1).

\item[6)] Consider the first entry of {\em Tape} (let's say it is
  node $i$).  The current vale of {\em FPTs} divided by $N-1$ is the
  first passage time $t_i$ of Eq.~(\ref{eq:mfpt_of_node}) relative to
  node $i$.\\ Store $t_i$ and the corresponding the node label $i$.
  Remove the first entry of {\em Tape} (but keep in memory the
  label $i$).  Decrease the {\em number-of-visits}$[i]$ and {\em
    L$^{old}$} by $1$.

\item[7)] If {\em number-of-visits}$[i]$ is equal to zero go to step
  8) otherwise go to step 9).

\item[8)] Decrease {\em counter} by $1$ and {\em FPTs} by
  $(N-1)$. Then go to step 5).

\item[9)] Select the value $L^*$ in the first entry of the array {\em
  not-first-passage}$[i]$. Set
\begin{equation*}
\centering
 {FPTs} = { FPTs} - (N-1) + (L^* - L^{old})
\end{equation*}
Remove the first entry of the array {\em not-first-passage}$[i]$. Go to step 5).
\end{itemize}

Steps $1-4$ perform the walk motion and add the sequence of visited
nodes in {\em Tape}. In step 2) we check if the node $j$ has not yet
been visited and if so in step 3) we store the first passage time
$t_{ij}=L-L^{old}$ in the variable {\em FPTs}.  When all nodes has
been visited at least once the algorithm enters in the loop $5-9$.
Steps $5-9$ repeatedly remove the entries at the beginning of {\em
  Tape} and compute, after each removal, the mean first passage time
$t_i$ of Eq. (\ref{eq:mfpt_of_node}) relative to each removed node
$i$. If a node label is no longer contained in {\em Tape} the
algorithm goes back to the $1-4$ loop until all nodes has been visited
at least once. The advantage of this strategy is that the estimated
mean first passage time $t_i$ for a certain node $i$ can be computed
using the mean first passage time $t_\ell$ of the node $\ell$ that
precedes the node $i$ in {\em Tape} as described by the recursive
equation in step 9). The numerical simulation is left running until
  the estimate of $t_i$ is averaged over 1000 realizations for each
  node $i$.

\begin{figure}[ht]
\centering
\includegraphics[width=2in]{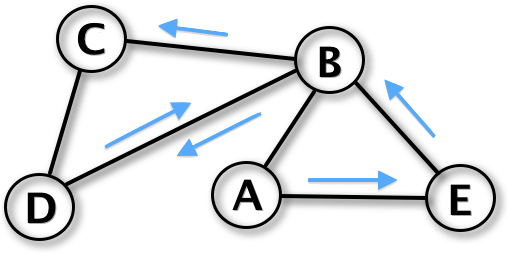}
\begin{center}
\begin{tabular}{|c|c|c|c|c|c|c|c|}
\hline
Node Sequence ({\em Tape}) & A  & E & B & D & B & C & $...$ \\
\hline
Time Passages ($L$)& 0 & 1 & 2 & 3 & 4 & 5 &$...$\\
\hline
\end{tabular}
\end{center}
\caption{ A walk on a network of $N=5$ nodes. The passage on node {\bf
    B} at time $L=2$ is a genuine first passage relative to the walk
  starting at nodes {\bf A} and {\bf E}. The passage on node {\bf B}
  at time $L=4$ is a genuine first passage only when the first three
  nodes are removed from {\em Tape} and we consider the walk starting
  on node {\bf D}.}
\label{fig:graphexample}
\end{figure}

To further clarify the key strategy used in the algorithm let us give an example on a small graph with $N=5$ nodes and a walker path illustrated in Fig. (\ref{fig:graphexample}).
The second passage on node {\bf B} at time $L=4$ is excluded in the
computation of $t_{{\bf A}}$ because the first passage on node {\bf B}
has already occurred at the second hop $(L=2)$.  However the value $L=4$ is
added at the end of the array {\em not-first-passage}$[{\bf B}]$ to be
used later (let's call this value $L^*=4$).  Indeed when the first
three entries of {\em Tape} are removed (loop $1-9$) and we consider the walk
starting on node {\bf D} the second passage in node {\bf B} occurred
at $L=4$ is now a genuine first passage.  At this time, because we
have removed three entries from {\em Tape}, we have $L^{old}=3$
and the correct number of hops between node {\bf D} and the first
passage on node {\bf B} is given by $L^*-L^{old}=4-3=1$.  The value
$L^*-L^{old}$ is then used in the computation of $t_{{\bf D}}$.

\begin{figure}
  \centering
  \includegraphics[scale=0.35]{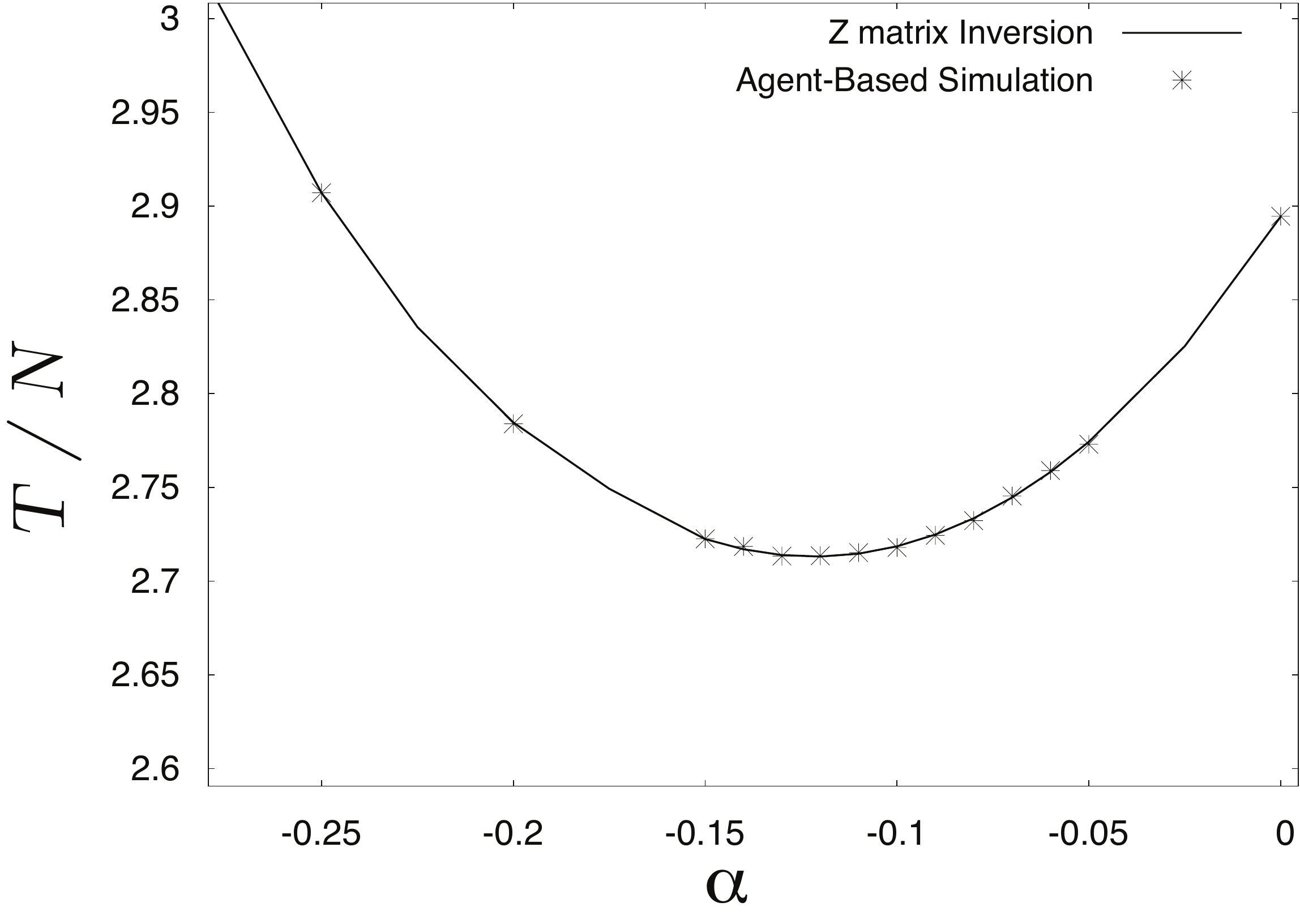}
  \caption{The value of the graph mean first passage time $T$
    normalized by the number of nodes $N$ as a function of $\alpha$,
    for the \mbox{InternetAS} network. The agent-based simulation
    (crosses) provides a very good approximation of the exact values
    obtained using the fundamental matrix $Z$ (solid line).}
\label{fig:validation}
\end{figure}
In Fig. \ref{fig:validation} we show a validation of our agent-based
simulation by comparing it with the result of the inversion of the $Z$
matrix for a small network.

%
%
%
 \begin{table*}
 \begin{tabular}{|c|c|c|c|c|c|c|c| }
 \hline
 \textbf{ Network } & \textbf{$Nodes$} & \textbf{$Edges$} & \textbf{$\langle
   k\rangle$}  &  \textbf{$ \nu$} & \textbf{$ \alpha_{\rm min} ^{R}$} & \textbf{$ \alpha_{\rm min}^{T}$} & \textbf{$ \alpha_{\rm min}^{C}$} \\
 \hline
 \hline
 \textbf{Synthetic Model:} & & & & &   & &\\
 \hline
 ER   & $10^4$  & $2  \cdot  10^4$  & $4 $   & $-$ & $-0.78$ & $-0.46  \pm 0.01$ & $-0.90$ \\
 ER   & $10^4$                   & $5 \cdot 10^4$    & $10$  & $-$ & $-0.89$ & $-0.78 \pm 0.01$ & $-1.00$  \\
 ER   & $10^4$                   & $2  \cdot 10^5$   & $40$  & $-$ & $-0.97$  & $-0.96 \pm 0.01$  & $-1.00$ \\
 \hline
Conf. Model ($\gamma=3$)   & $10^3$  & $4037 $ & $8$        & $-$ & $-0.87$ & $-0.65\pm 0.01$ & $-0.65 \pm 0.05$ \\
Conf. Model ($\gamma=3$)  & $5\cdot10^4$  & $ 21458  $ & $8$  & $-$  & $ -0.86$ & $-$  & $-$\\
Conf. Model ($\gamma=3$)   & $10^3$  & $8764 $ & $17.5$      & $-$ & $-0.94$ & $-0.87\pm 0.01$ & $-0.95 \pm 0.03$   \\
Conf. Model ($\gamma=3$)  & $10^3$  & $ 28522  $ & $57$  & $-$  & $-0.99$ & $-0.98\pm 0.01$  & $-0.98 \pm 0.02$\\
Conf. Model ($\gamma=2.5$)  & $10^4$  & $ 232722 $ & $46.5$  & $-$  & $-0.98$ & $-0.98\pm 0.01$&$ -0.98 \pm 0.02$ \\
 \hline
 BA(m=3)   & $10^3$  & $3 \cdot 10^3$ & $5.9$ & $-$  & $-0.78$ & $-0.36 \pm 0.02$  & $-0.30 \pm 0.05$ \\
 BA(m=5)   & $10^3$  & $5 \cdot 10^3$ & $9.9$ & $-$  & $-0.98$ & $-0.67\pm 0.01$ & $-0.50 \pm 0.05$ \\
 BA(m=20) & $10^3$  & $2 \cdot 10^4$ & $40$  & $-$  & $-1.02$ & $-0.99\pm 0.01$ & $-1.05 \pm 0.05$ \\
 \hline
 BA(m=3)   & $10^5$  & $3 \cdot 10^5 $ & $6$        & $-$ & $-0.76$ & $-$ &$-$ \\
 BA(m=5)   & $10^5$  & $5 \cdot 10^5 $ & $10$      & $-$ & $-0.99$ & $-$ &$-$ \\
 BA(m=20) & $10^5$  & $2 \cdot 10^6 $ & $39.9$  & $-$  & $-1.02$ & $-$  &$-$ \\
 \hline
 BA(m=20) & $4 \cdot 10^4$  &  $8 \cdot 10^5$ & $39.9$  & $-$ & $-1.02$ & $-$  &$-$ \\
 \hline
 \hline
 \textbf{Real-world networks:} & & & & & & & \\
 \hline
 Gnutella(P2P) ~\cite{Ripeanu2002}& 62561 & 147877   & 4.72  & $-$ &  $-0.91$  & $-0.55 \pm 0.02$  & $-0.5 \pm 0.05$ \\
 PairsFSG \cite{Nelson1998}& $10618$ & $63787$   & $12.01$  & $-$ & $-0.89$ &  $-0.68 \pm 0.01$ &  $-0.60 \pm 0.05$\\ 
 Email URV ~\cite{Guimera2003}& 1133 & 5451 &  $9.62$   & $0.05$ & $-0.76$ &$-0.62\pm 0.01$ & $-0.70 $\\
 Jazz  ~\cite{Gleiser2003}& 198 & 2742          & $ 29.01$ & $0.11$ & $-0.70$ & $-0.70\pm 0.01$ & $-0.90 $\\
 amazon ~\cite{Leskovec2007b}& 410236 & 2439437   & 11.89  & $-$ &  $-0.68$  & $-$  &$-$ \\
 USPower~\cite{Watts1998} & 4941 & 6593  & 2.66 & $-0.02$ &  $-0.66$  &  $ 0.17 \pm 0.01$ &  $-0.12 \pm 0.05$\\
 SCN ~\cite{Newman2001}& 12722 & 39967  & $6.28$     & $ 0.18$ & $-0.64$ & $-0.32\pm 0.01$  & $-0.5  \pm 0.1$\\ 
 ca-CondMath ~\cite{Newman2001} & 21363 & 91286  & 8.54  & $0.16$ &  $-0.63$ & $-0.43 \pm 0.01$    & $-0.47 \pm 0.05$ \\
 ca-HepTh ~\cite{Newman2001}& 8638 & 24806  & $5.76$  & $0.19$ &  $-0.65$ &  $-0.37 \pm 0.01 $  &  $-0.47 \pm 0.05$ \\
 ca-AstroPh ~\cite{Newman2001}& 17903 & 196972  & $22.00$  & $0.22$ & $ -0.62 $ &  $-0.58\pm0.01$  &$ -0.52 \pm 0.05$ \\
 ca-ASTRO ~\cite{Newman2001}& 13259 & 123838   & 18.68  & $0.34$ & $ -0.59$ & $-0.54 \pm0.01$   & $-0.60 \pm 0.05$ \\
 ca-HepPh ~\cite{Leskovec2007a}& 11204 & 117619  & $20.99$  & $0.54$ &  $-0.57$  &  $-0.51 \pm 0.01$ &   $-0.43 \pm 0.05$ \\
 pgp ~\cite{Boguna2004}& 10680 & 24316   & $4.55$  & $-$ &  $-0.48$  & $-0.20 \pm 0.02$   & $-0.25 \pm 0.05$ \\
 \hline
 C.Elegans ~\cite{Watts1998} &  279 & 2287 & $16.39$ & $-0.15$ & $-0.79$ & $-0.68\pm 0.01$ & $-0.7 \pm 0.1$ \\
 bio-Yeast ~\cite{Sun2003}& 2312 & 7165   & 6.20  & $-0.42$ &  $-0.44$ & $-0.32 \pm 0.01$ &  $-0.20 \pm 0.02$ \\
 www-Google ~\cite{Leskovec2009}& 855802 & 4291352   & 10.03 & $-0.42$ & $ -0.43 $ & $-0.33 \pm 0.02$  &$-0.25 \pm 0.1$  \\
 soc-Slashdot ~\cite{Leskovec2009}& 82168 & 582290   & 14.17  & $-0.78$ &  $-0.43$ & $-0.38 \pm 0.02$  & $-0.16 \pm 0.06$\\
 soc-Epinions ~\cite{Richardson2003} & 75877 & 405739   & 10.69  & $-$ & $ -0.39$ & $-0.34 \pm 0.02 $ &  $-0.24 \pm 0.01 $ \\
 Actors  ~\cite{Watts1998}& 374511 & 1222908   & 6.53  & $-0.23$ & $ -0.37$  & $-0.35 \pm 0.02$  & $-0.28 \pm 0.06$ \\
 wordnet ~\cite{Fellbaum1998} & 75609 & 120473   & 3.18  & $-0.41$ &  $-0.30$  & $-0.11 \pm 0.02$  & $-0.20 \pm 0.05 $\\
 www-NotreDame ~\cite{Barabasi1999}& 325729 & 1090108   & $6.69$  & $-0.84$ &  $-0.29$ & $-0.1 \pm 0.05$  & $-$ \\
 www-Stanford ~\cite{Leskovec2009}& 255265 & 1941926   & 15.21  & $-0.72$ &  $-0.23$ & $-0.23 \pm 0.07$ & $-0.25 \pm 0.08$ \\
 www-BerkStan ~\cite{Leskovec2009}& 654782 & 6581870   & 20.10  & $-0.84$ &  $-0.25$ & $-0.29 \pm 0.05$ & $-$  \\
 caida ~\cite{Leskovec2007a}& 26475 & 53381     &  $4.03$   & $-0.50$ & $-0.15$ &$-0.12 \pm 0.01$ & $-0.15 \pm 0.05$\\
 InternetAS ~\cite{Pastor-Satorras2001}& 11174 & 23409  & 4.19  & $-0.52$ & $-0.14$ & $-0.12 \pm 0.01 $ &  $-0.11 \pm 0.01$ \\
 \hline
 USairport ~\cite{Opsahl2011}& 1572 & 17214   & $21.90$  & $-$ & $-0.58$ & $-0.55 \pm 0.01$ & $ -0.6 \pm 0.01$ \\  
 USairports500 ~\cite{Colizza2007} & 500 & 2980     & $11.92$  & $-$ & $-0.50$ & $-0.42$ & $-0.40$ \\
 netscience.net ~\cite{Newman2006} & 379 & 914  &  $4.82$  & $-$ & $-0.67 $& $-0.58$ & $-0.20 $\\
\hline
 \end{tabular}
 \caption {Values of $\nu$ and of $\alpha_{\rm min}$ for MRT, MFPT and
   MCT in synthetic and real-world complex networks. The mean-field
   approximation gives correct results for synthetic uncorrelated
   networks (i.e., \Erdos--R\'enyi, configuration model and
   Barab\'asi-Albert networks) with sufficiently large values of
   $\langle k \rangle$ and $N$. The values of $\nu$ are missing for
   those networks for which $k_{nn}(k)$ is not a power-law.}
 \label{tab:list_net}
 \end{table*}


\begin{thebibliography}{99}
\bibitem{note_cutoff} Notice that the starting configuration model
  network used here has a maximum degree $k_{max}=300$, which is
  larger than the structural cut-off of the network, as often observed
  in real-world networks.
\end{thebibliography}
\end{document}